\documentclass[aps,prx,twocolumn,superscriptaddress,showpacs,floatfix,longbibliography]{revtex4-2}
\usepackage{amsmath,amssymb,wasysym,graphicx}
\usepackage{times}
\usepackage{tikz}
\usepackage[varg]{txfonts}
\usepackage{textcomp}
\usepackage{tabu}
\usepackage{color}
\usepackage{xcolor}
\usepackage[colorlinks=true,citecolor=blue,urlcolor=blue,linkcolor=blue,hyperindex]{hyperref}
\usepackage{braket}
\usepackage{overpic}
\usepackage{booktabs}
\usepackage{tabularx}
\usepackage{makecell}

\usepackage[normalem]{ulem}
\usepackage{verbatim}
\usepackage{float}
\usepackage{color}

\usepackage{bm}
\DeclareMathOperator{\Tr}{Tr}

\pdfoutput=1
\definecolor{LinkColor}{rgb}{0.256,0.439,0.588}
\usepackage{hyperref}
\hypersetup{
colorlinks=true,
citecolor=LinkColor,
linkcolor=LinkColor,
urlcolor=LinkColor
}

\usepackage{multirow}

\begin{document}

\title{Precise computation of universal corner entanglement entropy at 2+1 dimension: \\ From Ising to Gaussian quantum critical points}

\author{Ben Lee-Yeung Ngai}
\thanks{These authors contributed equally}
\affiliation{Department of Physics, HK Institute of Quantum Science \& Technology and State Key Laboratory of Optical Quantum Materials, The University of Hong Kong, Pokfulam Road, Hong Kong SAR, China}

\author{Justin Tim-Lok Chau}
\thanks{These authors contributed equally}
\affiliation{Department of Physics, HK Institute of Quantum Science \& Technology and State Key Laboratory of Optical Quantum Materials, The University of Hong Kong, Pokfulam Road, Hong Kong SAR, China}

\author{Junchen Rong}
\affiliation{CPHT, CNRS, Ecole Polytechnique, Institut Polytechnique de Paris, Palaiseau, France}

\author{Meng Cheng}
\affiliation{Department of Physics, Yale University, New Haven, CT 06511, USA}

\author{Yuan Da Liao}
\email{ydliao@hku.hk}
\affiliation{Department of Physics, HK Institute of Quantum Science \& Technology and State Key Laboratory of Optical Quantum Materials, The University of Hong Kong, Pokfulam Road, Hong Kong SAR, China}

\author{Zi Yang Meng}
\email{zymeng@hku.hk}
\affiliation{Department of Physics, HK Institute of Quantum Science \& Technology and State Key Laboratory of Optical Quantum Materials, The University of Hong Kong, Pokfulam Road, Hong Kong SAR, China}

\date{\today}

\begin{abstract}
Computing the subleading logarithmic term in the entanglement entropy (EE) of (2+1)d quantum many-body systems remains a significant challenge, despite its central role in revealing universal information about quantum states and quantum critical points (QCPs). Building on recent algorithmic advances that enable the stable calculation of EE as an exponential observable~\cite{zhouIncremental2024,zhangIntegral2024,liaoExtracting2024}, we develop a {\it bubble basis} projector quantum Monte Carlo (QMC) algorithm to precisely and efficiently compute the universal corner of EE at QCPs in a (2+1)d square-lattice transverse-field Ising model augmented with a four-body interaction.  Turning on this interaction allows us to trace an Ising critical line, reaching the tricritical point, and then a line of first-order phase transition. In (2+1)d, the tricritical point is described by the Gaussian theory, where a theoretical calculation of the corner logarithmic term in the 2nd R\'enyi entropy term is available~\cite{UniversalCasini2007}. 
Our QMC results are in quantitative agreement with this theoretical value, providing a highly nontrivial benchmark of the algorithm.
Furthermore, we also study the R\'enyi EE at the Ising critical line and on the first-order transition line, obtaining results consistent with theoretical expectations.
These findings establish the long-sought connection between the universal values of an exactly solvable limit and those of a strongly correlated regime at (2+1)d.

\end{abstract}

\maketitle

\section{Introduction}
\label{sec:intro}
Entanglement is the quintessential property of quantum mechanics~\cite{einstein1935can, bell1964einstein, clauser1969proposed}. As the investigation of quantum many-body states of matter advances, precise computation and quantification of the quantum entanglement therein become an urgent task~\cite{vedral2008quantifying,amico2008entanglement}. Among many available entanglement measures, the entanglement entropy (EE) plays a central role, in that, it has greatly deepened our understanding and refined the classification of quantum states of matter~\cite{laflorencie2016quantum,isakovTopological2011,DEmidioEntanglement2020,zhaoScaling2022,songExtracting2024,DEmidioUniversal2024,liaoUniversal2025,songEvolution2025}. For example,  in (1+1)d systems, the sublinear scaling of EE with subsystem size underpins the efficiency of matrix product state representations~\cite{ciracMatrix2021}.
In (2+1)d, universal contribution to EE was first identified in exactly solvable models~\cite{levinDetecting2006,kitaevTopological2006}, and later measured in prototypical interacting models using quantum Monte Carlo (QMC) methods~\cite{isakovTopological2011,zhaoMeasuring2022}. Overall, EE enables direct probes of long-range entanglement, the structure of topological order, and the nature of the emergent fractionalised degrees of freedom.

However, a fully generic EE computation scheme for (2+1)d quantum many-body systems remains a major challenge, primarily due to the exponential complexity in accessing the wave function. An important feature of EE in (2+1)d quantum critical states is the universal corner term, i.e. the subleading logarithmic contribution to the EE.  To date, analytical results for this universal term have been obtained only in free theories at (2+1)d~\cite{UniversalCasini2007,UniversalHelmes2016}, for both R\'enyi and von Neumann EEs. For interacting theories, analytically controlled calculations have only been performed for O($N$) critical point in the large $N$ approximation for the von Neumann entropy \cite{Whitsitt:2016irx}. 

Over the years, numerous efforts have been made in developing QMC algorithm for high-precision computation of R\'enyi EE to extract the universal corner term at the (2+1)d Ising and O($N$) critical points~\cite{InglisEntanglement2013,stoudenmireCorner2014,CornerKallin2014,EntanglementHelmes2014,songExtracting2024,liaoExtracting2024}. Despite the progress, the current state of the art is still not fully satisfactory, for the following reasons 

\begin{itemize}
\item The precise computation of R\'enyi EE in QMC for (2+1)d systems is hampered by the lack of efficient algorithms, a challenge rooted in the exponential nature of the entanglement observable itself~\cite{zhangIntegral2024,zhouIncremental2024,liaoExtracting2024}, such that in a naive computation scheme of the ratio of replica partition functions, the coefficient of variation of obtained EE will increase exponentially with the system sizes~\cite{panStable2023,liaoUniversal2025}, rendering the computation complexity grows exponentially if one wants to control the relative percentage of standard deviation over the mean value.

\item These universal terms are subleading to the dominant area-law contribution that scales extensively with the subsystem size. Isolating the subleading contribution directly from the finite-size EE data often suffers from substantial error bars,  limiting their numerical precision. 

\item There has been no attempt to perform a benchmark study against the exactly solvable results for free theories and then compute EE across the interacting (2+1)d critical points within one set of QMC simulations. Such a setup would allow one to first quantitatively validate the QMC algorithm on reproducing the free-theory results, and only then place further confidence in the extraction of universal corner terms for the interacting quantum critical points, including a quantitative assessment of their distance from the Gaussian (free-theory) results.
\end{itemize}
It is due to these issues that the overall status of the precise computation of the universal corner coefficient for interacting QCPs remains somewhat ambiguous.

\begin{figure*}[htp!]
\includegraphics[width=\textwidth]{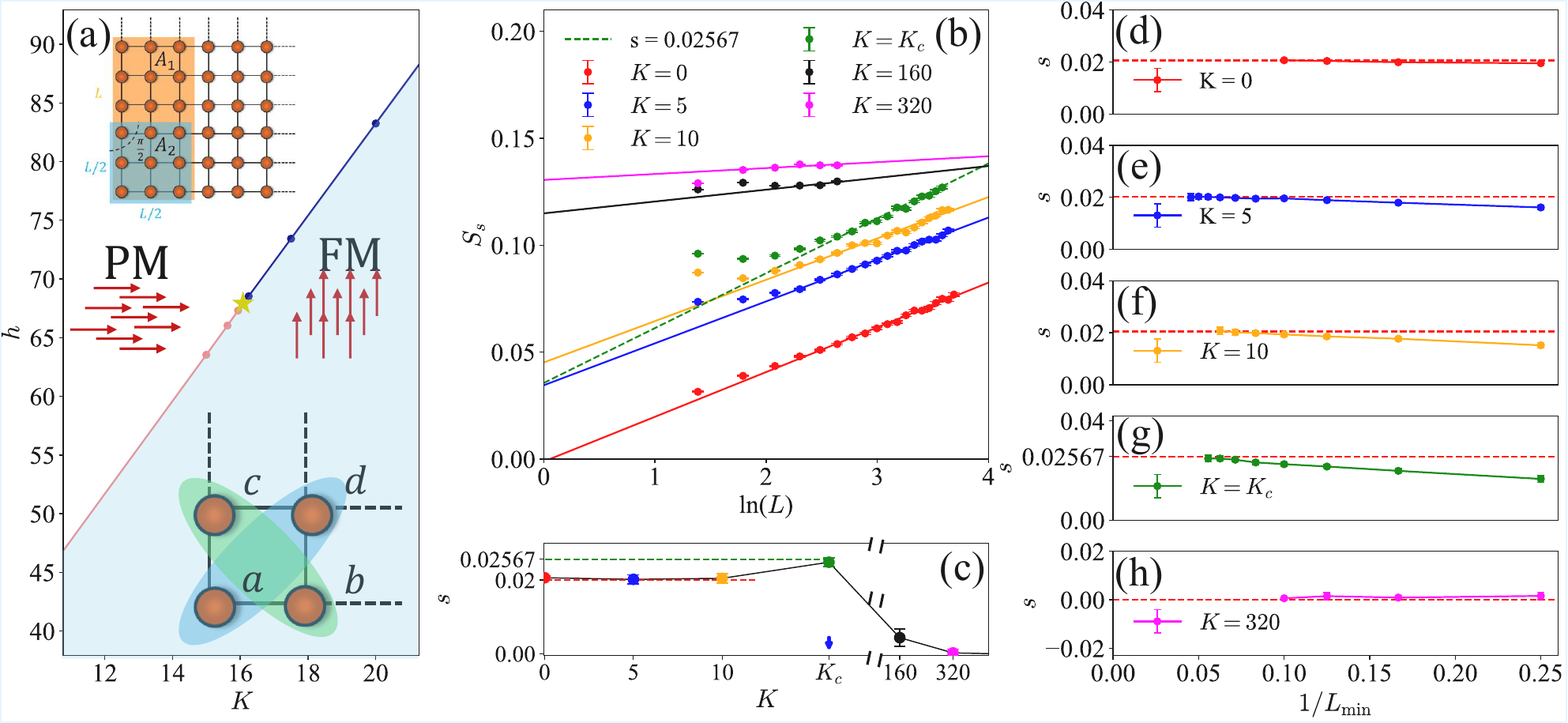}
\label{fig:fig1}
\caption{\textbf{Model, phase diagram and the universal corner EE from Ising to Gaussian QCPs.} 
(a) Phase transition on $h-K$ plane. The light pink line represents the continuous phase transition, while the blue line indicates the first-order phase transition. The dots indicate the points where we performed the crossing-point analysis to determine the locations of the phase transitions. The PQMC determined Gaussian fixed point is identified as the yellow star. The top left inset in (a) illustrates the entanglement regions. In a $L \times L$ square lattice, the entangled region $A_1$ has the dimension $L \times L/2$ (orange background) and it has a smooth boundary [Eq. \eqref{eq:Second order REE of smooth boundary}]. The region $A_2$ has a $L/2 \times L/2$ (blue background) and it has four 90$^\circ$ corners log-contribution [Eq. \eqref{eq:Second order REE of corner boundary}]. The lattice is periodic along both directions (denoted by the dotted lines), and the entanglement regions $A_1$ and $A_2$ have the same boundary length. The bottom right inset in (a) illustrates the action of $K$-term in the Hamiltonian of Eq.~\eqref{eq:eq1}. 
(b) Scaling of the subtracted entanglement entropy $S_s$. The dependence of $S_s$ on $\ln(L)$ at the phase transition boundary for selected values of $K = 0$, 5, 10, $K_c$, 160 and 320 are shown, where $K_c = 16.02(6)$ is the Gaussian QCP. 
(c) The universal log-coefficient $s$, extracted from the data in (b), as a function of $K$. 
(d), (e), and (f) show the extrapolation for $K=0$, $K=5$, and $K=10$ at their corresponding (2+1)d Ising QCPs. The computed corner log-coefficients converge to a universal average value of $s=0.020(1)$ as $L_{\min}$ increases. 
(g) corresponds to the Gaussian fixed point at $K_c$, where the corner log-coefficient converges to the universal value of $s=0.025(1)$.
(h) is for $K=320$ at its first-order phase transition point; here, the computed corner log-coefficients vanishes as $s=0.000(1)$.}
\end{figure*}

In this work, we address these issues with advancements in three aspects.
First, we develop an advanced projector QMC (PQMC) framework based on a {\it bubble basis}. Its key innovation lies in vastly improving the efficiency of the incremental SWAP algorithm~\cite{liaoExtracting2024,zhouIncremental2024} for EE calculation. 
While the conventional $\sigma^z$-basis approach scales as $O(m^2)$ with projection length $m$~\cite{InglisEntanglement2013}, our technique reduces the complexity to $O(mP)$, where $P\ll m$ at critical points, enabling a superior scaling for high-precision calculations for EE. 
The acceleration originates from how the SWAP operator is sampled within the incremental framework in the two bases. In the $\sigma^z$ basis, sampling the SWAP operator when updating a configuration at a given projector slice requires evolving the state to the central projector slice, leading to an overall quadratic scaling. In contrast, within the bubble basis, the sampling can be performed with $O(P)$ cost at any projector slice, where $P$ is a slice-independent parameter. This fundamental distinction improves the total scaling from $O(m^2)$ to $O(mP)$.

Second, we successfully integrate the subtracted corner entanglement entropy technique~\cite{liaoExtracting2024} into the bubble basis PQMC. The combination of these methods allows the dominant area-law term in the EE to be directly eliminated during the QMC sampling process. As a result, the coveted subleading logarithmic term becomes the leading contribution in the simulation output, enabling a precise and direct extraction of the universal corner coefficient.  

Third, we design a (2+1)d square lattice transverse-field Ising model (TFIM) augmented with a four-body interaction term. Tuning the relative strength between the two-spin and four-spin couplings allows us to trace an Ising critical line that ends on a tricritical point beyond which the transition becomes first order. The tricritical point is a conformal field theory with two relevant operators. In (2+1)d, this is precisely the Gaussian theory,
where a theoretical calculation of the corner logarithmic term in the 2nd R\'enyi EE term is available~\cite{UniversalCasini2007,UniversalHelmes2016}.

With these three advancements, we apply our {\it bubble basis} incremental SWAP algorithm along this path, and successfully extract the universal corner contribution to the EE. At the Gaussian fixed point, the resulting logarithmic coefficient for four 90$^\circ$ corners agrees perfectly with the analytical value of $s = 0.02567$~\cite{UniversalCasini2007} -- our QMC estimate is $s = 0.025(1)$ -- the number in parentheses indicates the uncertainty in the last significant digit. This value is clearly distinguishable from the Ising universal coefficient obtained along the same path, $s = 0.020(1)$ for four 90$^\circ$ corners. The accurate determination of these universal terms across the (2+1)d QCPs -- from the Ising to the Gaussian universality class -- demonstrates the reliability of our approach in capturing challenging entanglement properties. It provides a high-precision benchmark for future EE computations and establishes a long-missing link between exactly solvable limits and strongly correlated regimes in (2+1) dimensions.

\section{Results}
\label{sec:results} 
We directly start with explaining the main results of the paper: our model design (Sec.~\ref{sec:resultsA} ), and then our PQMC results of the phase diagram (Sec.~\ref{sec:resultsB}) and the computed 2nd order R\'enyi EE reults along the parameter path from Ising to Gaussian QCPs, to monitor the changes of the universal corner log-coefficients (Sec.~\ref{sec:resultsC}). The detailed explanation of the analytic computation of EE at the Gaussian fixed point and our algorithmic advancements of {\it bubble basis} PQMC and the superior EE computation therein, will be given in Sec.~\ref{sec:method}.

\subsection{From Ising to Gaussian at (2+1)d}
\label{sec:resultsA} 
The Hamiltonian of a TFIM on a square lattice, augmented with a four-body $K$ interaction term, is given as
\begin{equation}
    H = -J\sum_{\langle i,j\rangle} \sigma^z_i \sigma^z_j - h \sum_i \sigma^x_i -K \sum_{(a,d)(b,c)} (\sigma_a^z\sigma_d^z+\mathbf{I}) (\sigma_b^z\sigma_c^z+\mathbf{I}),
\label{eq:eq1}
\end{equation}
where $\bm{\sigma}=\{\sigma^x,\sigma^y,\sigma^z\}$ denotes the Pauli matrices,  $i$ labels a lattice site, and  $\langle i,j\rangle$ denotes a nearest-neighbor pair of sites. 
The system has $N=L\times L$ sites in total, where $L$ is the linear size. We consider the ferromagnetic case, set $J=1$ as the energy unit, and take $h$ to be the transverse-field strength. To realize the Gaussian fixed point, we consider a  $K$-term into the TFIM
where $(a,d)$ and $(b,c)$ denote two distinct bonds in a plaquette, as illustrated in the lower right inset of Fig.~\ref{fig:fig1} (a). 

According to the RG flow analysis provided in the Appendix~\ref{appA}, the model can be effectively described by a quantum field theory of a single scalar field, with the potential $V=\frac{m^2}{2}\phi^2+\frac{1}{4!}\lambda_4 \phi^4 +\cdots$. 
At (2+1)d, the lattice coupling $h/J$ and $K/J$ controls the value of $m^2$ and $\lambda_4$-term in the field theory. 
When $\lambda_4 > 0$, going from $m^2<0$ to $m^2>0$, the system experiences a (2+1)d Ising QCP from ferromagnetic (FM) to paramagnetic (PM) phases. 
When $\lambda_4<0$, the phase transition becomes first order.
The fine-tuned point of $\lambda_4=0$ will give rise to a tricritical point.
This is also known as the tricritical Ising CFT at its upper critical dimension $d=3$, and examples of real-world systems with tricritical Ising fixed-points are ${}^3$He–${}^4$He mixtures~\cite{blumeIsing1971,maciolekPhase2004,farahmandPhase2015}, four-component fluid mixtures near room temperature~\cite{langEquilibrium1975} and metamagnets~\cite{nelsonRenormalization1975,fisherUniversality1975}. At the upper critical dimension, the tricritical Ising CFT is then described by the Gaussian fixed point (free theory). 
The CFT has two relevant operators $\phi^2$ and $\phi^4$; we have to tune the coupling constants of both operators to zero to reach the critical point. 
This explains why it can only appear as a single point in the two-dimensional phase diagram in Fig.~\ref{fig:fig1} (a). 
The next operator, $\phi^6$, is marginally irrelevant, yielding logarithmic finite-size corrections to critical exponents.
Previous Monte Carlo simulations tried to detect the logarithmic corrections, but their status does not appear to be definite~\cite{maciolekPhase2004}.

Of course, the mapping from the field theory parameter $m^2$ and $\lambda_4$ is not known a priori. We perform an unbiased QMC simulation and obtain the phase diagram in Fig.~\ref{fig:fig1} (a), which is fully consistent with the field theory result, as we now turn to.

\begin{figure*}[htp!]
\includegraphics[width=0.8\textwidth]{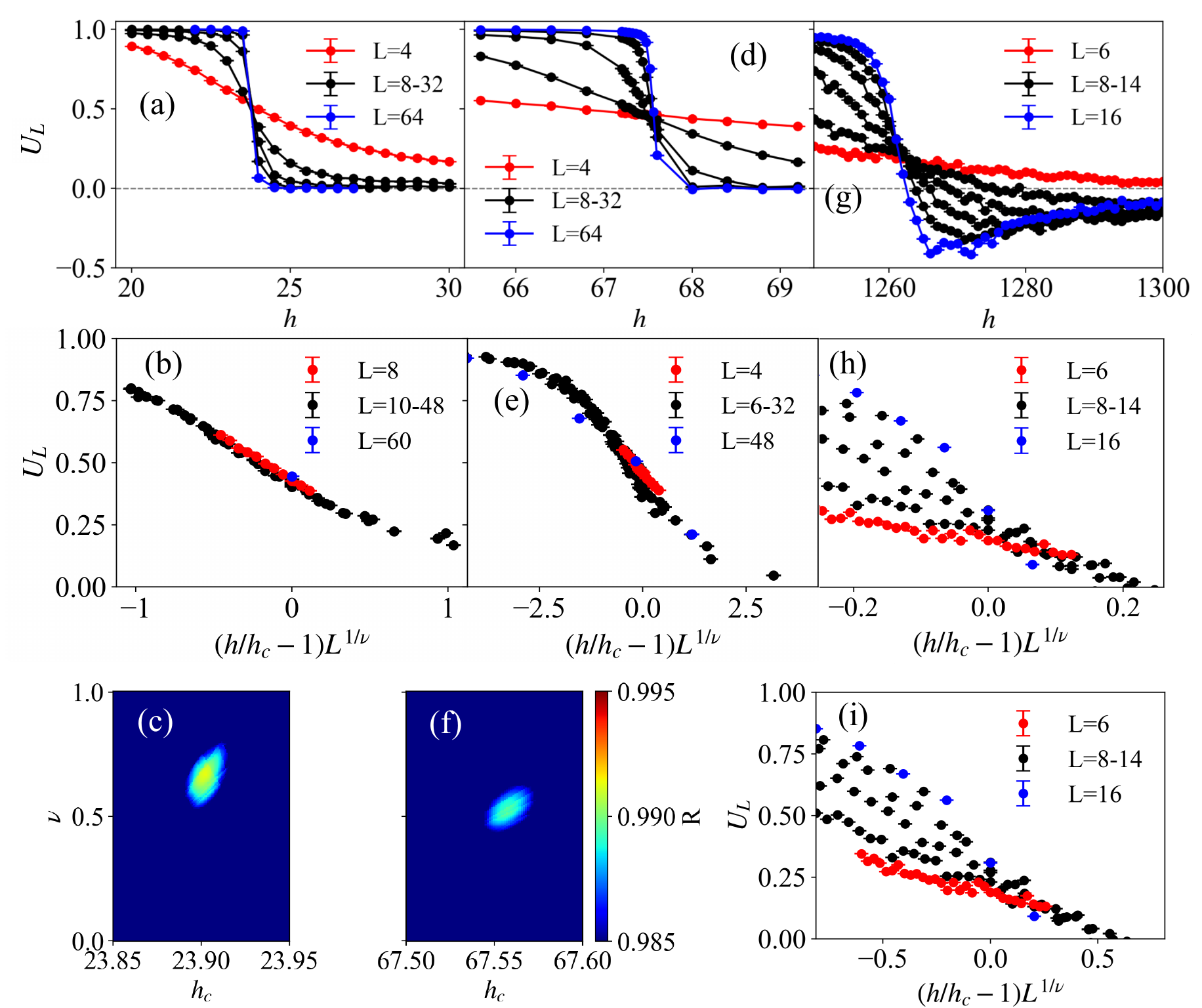}
\caption{\textbf{Finite-size scaling analysis of the phase transitions driven by transverse field $h$ at different values of parameter $K$.} 
(a-c) $K = 5$: (a) Binder cumulant $U_L$ as a function of $h$ for different system sizes up to $L = 64$. The crossing point of the curves determines the critical field $h_c^{K=5}=23.9(5)$.
(b) Data collapse of $U_L$ using the finite-size scaling relation $U_L \sim (h/h_c - 1)L^{1/\nu}$ with the 3D Ising universality critical exponent $\nu = 0.63$. The high quality of the collapse confirms a continuous phase transition.
(c) Heat map of the $R$ statistic for the data collapse, evaluated over a range of $\nu$ and $h_c$ values. The optimal region of high $R$ aligns with $\nu = 0.63$, providing further evidence that the continuous phase transition belongs to the 3D Ising universality.
(d-f) $K = 16.02$: (d) Binder cumulant $U_L$ versus $h$ for different $L$ up to 64. The intersection point identifies $h_c^{K=16.02}=67.57(4)$.
(e) Data collapse of $U_L$ using the Gaussian (free) universality critical exponent $\nu = 0.5$, indicating a continuous phase transition belonging to this class.
(f) Heat map based on the $R$ criterion for the finite-size scaling relation, exploring different values of $\nu$ and $h_c$. The bright region confirming the best data collapse includes $\nu = 0.5$, consistent with the Gaussian universality.
(g-i) $K = 320$: (g) Binder cumulant $U_L$ as a function of $h$ for different $L$ up to 16. The emergence of negative $U_L$ values for $L=12$ and $L=16$ at certain values of $h$ is a signature of a first-order phase transition.
(h) Attempted data collapse of $U_L$ using the 3D Ising critical exponent ($\nu = 0.63$) fails.
(i) Attempted data collapse using the Gaussian critical exponent ($\nu = 0.5$) also fails. The poor quality of both collapse attempts provides additional evidence for a first-order phase transition at this parameter.}
\label{fig:fig8}
\end{figure*}

\subsection{Phase Diagram}
\label{sec:resultsB}
Fig.~\ref{fig:fig8} (a), (d) and (g) show the analysis of the Binder cumulant 
$U_L=\frac{3}{2}\left(1-\frac{\left\langle M_z^4\right\rangle}{3\left\langle M_z^2\right\rangle^2}\right)$ for various values of the tuning parameter $h$ at $K = 5$, $K = 16.02$, and $K = 320$, with the linear system sizes from $L=4$ up to $L = 64$. The Binder cumulants approach 1 at small $h$ and $0$ at large $h$ for all three $K$ values, indicating a transition from a FM to a PM phase with increasing $h$. The nature of this transition, however, depends on the value of $K$. For $K = 5$ and $K = 16.02$, the Binder cumulant exhibits behavior characteristic of a continuous phase transition. In contrast, for $K = 320$, the transition is first-order, as evidenced by the emergence of negative values in the Binder cumulant~\cite{binderMonte1992,vollmayrFinite1993} to the right of the critical point for $L > 10$.

To further classify the universality classes of the continuous transitions, we performed a finite-size scaling analysis. For the continuous transition at $K = 5$, the data collapse was successfully achieved using the (2+1)d Ising critical exponent (correlation length exponent $\nu=0.63$), as illustrated in Fig.~\ref{fig:fig8} (b), confirming that it belongs to the Ising universality class and give rise $h_c=23.9(5)$.
For $K = 16.02$, a satisfactory data collapse was obtained only when using Gaussian exponent (correlation length exponent $\nu=1/2$), shown in Fig.~\ref{fig:fig8} (e), indicating its idenfication as the Gaussian universality class, and the corresponding $h_c=67.57(4)$. For the first-order transition at $K = 320$, attempts to collapse the data using either (2+1)d Ising or Gaussian exponents were unsuccessful (Fig.~\ref{fig:fig8} (h) and (i)), which provides additional support for the first-order nature of the transition at this point.
The very large value of $K$ was selected specifically to distance the system from the Gaussian critical point, thereby cleanly isolating and demonstrating the characteristic first-order behavior.

We also set the position of the QCP ($h_c$) and the correlation length exponent ($\nu$) as free parameters and record the coefficient of determination, $R^2$, of the data collapse with each set of ($h_c$, $\nu$) such that the optimized collapse can be determined with the optimized choice of the ($h_c$,$\nu$) for different $K$ cuts. The corresponding heatmap are shown in Fig.~\ref{fig:fig8} (c) and (f). It is obvious from these analyses that for $K=5$, the optimized data collapse can be obtained only when the $\nu=0.63(5)$ which correspond to the (2+1)d Ising QCP, whereas for $K=16.02$, the optimized data collapse can be only be obtained when $\nu=0.52(5)$, which correspond to the Gaussian QCP.

The raw data presented in Fig.~\ref{fig:fig8} were acquired using the PQMC method in the $\sigma_z$ basis. This choice was made because, as will be discussed in the Secs.~\ref{sec:methodB} and \ref{sec:methodC}, the $\sigma_z$ basis PQMC demonstrates better efficiency than bubble basis PQMC for calculating standard physical observables such as energy, magnetization, and the Binder cumulant, but worse efficiency in computing the exponential observables~\cite{zhangIntegral2024}, such as EE and free energy.

\subsection{Universal terms in the EE}
\label{sec:resultsC}
We thus compute the 2nd order R\'enyi EE $S_2^A$, which is defined as 
\begin{equation}
    S_2^A = -\ln(\Tr \rho_A^2),
\end{equation}
where $\rho_A$ is the reduced density matrix for the region $A$.
We will refer it as EE for short throughout the paper unless otherwise stated. For all phases that we are interested in, EE is known to obey the following scaling for a sufficiently large $A$:
\begin{equation}\label{eq:S2scaling}
    S_2^A = a l_A - s \ln(l_A) + c,
\end{equation}
where $ l_A $ is the boundary length of the entanglement region (typically proportional to the linear system size $ L $, as illustrated in the inset of Fig.~\ref{fig:fig1} (a)), $ a $ is a non-universal coefficient associated with the area-law term, $ s $ is a universal coefficient quantifying the subleading logarithmic correction, which depends on the presence of sharp corners in the entanglement region and the ground state properties, and $ c $ is a constant. The coefficient $s$ is vital for characterizing quantum criticalities and their underlying conformal field theories.

At the Gaussian fixed point, the logarithmic correction arising from sharp corners is well-established and considered exactly solvable, as discussed in Sec.~\ref{sec:methodA}. For the entanglement region shown in the inset of Fig.~\ref{fig:fig1} (a), which contains four 90$^\circ$ corners, the universal coefficient of this logarithmic correction at the Gaussian point is known to be $s=0.02567$~\cite{UniversalCasini2007,UniversalHelmes2016} (cf. Fig.~\ref{fig:theory}).

We employ the bubble basis PQMC method to compute the EE of the TFIM, enhanced further by the subtracted corner entanglement entropy (SCEE) technique~\cite{liaoExtracting2024,songExtracting2024,songEvolution2025}. The SCEE technique is specifically designed to isolate the universal log-coefficients originating from sharp corners on the entanglement boundary during one QMC simulation.  This is achieved by calculating the difference between the EEs of two strategically chosen subregions, $ A_1 $ and $ A_2 $, within the same system, as depicted in the upper inset of Fig.~\ref{fig:fig1} (a). Subregion $ A_1 $ possesses a smooth boundary, and its EE follows the form
\begin{equation}
    S_2^{A_1} = a l_{A_1} + \gamma_1,
    \label{eq:Second order REE of smooth boundary}
\end{equation}
while subregion $ A_2 $, which contains corners, exhibits an EE given by
\begin{equation}
    S_2^{A_2} = a l_{A_2} - s \ln l_{A_2} + \gamma_2 .
    \label{eq:Second order REE of corner boundary}
\end{equation}
The subregions are selected such that their boundary lengths are equal: $ l_{A_1} = l_{A_2} = 2L $. The SCEE is then defined as the difference between $  S^{A_1}_{2} $ and $ S^{A_2}_{2} $:
\begin{equation}
    S_s = S^{A_1}_{2} - S^{A_2}_{2} = s \ln L + \gamma,
    \label{eq:SCEE definition}
\end{equation}
where $ \gamma = \gamma_1 - \gamma_2 $.
This differential approach inherently cancels the dominant, non-universal area-law term in the entanglement entropy. 
Consequently, the universal corner contribution emerges as the leading term in the scaling analysis, enabling its precise and straightforward extraction.
Moreover, the quantity $S_s$ can be obtained directly in the QMC simulation. This approach is more efficient and yields better precision than first calculating $S_2^{A_1}$ and $S_2^{A_2}$ separately and then subtracting them, as it avoids the accumulation of errors from the subtraction operation, which was performed in previous QMC works~\cite{InglisEntanglement2013,stoudenmireCorner2014,CornerKallin2014,EntanglementHelmes2014}. 

Fig.~\ref{fig:fig1} (b) demonstrate the $S_s$ versus $\ln(L)$ along the critical boundary in Fig.~\ref{fig:fig1} (a) for various $K$ values. $L_{\max}$ takes values of 40, 38, 38, 36, 14, and 14 for $K = 0$, 5, 10, $K_c$, 160, and 320, respectively. We found as $L$ increases, the expected $S_s \sim s\ln(L)$ behavior (Eq.~\eqref{eq:SCEE definition}) manifests, and for $K=0,5,10<K_c$, i.e., the (2+1)d Ising QCPs, the curves are all parallel (although their intercepts are different) which implies the equal value of the universal log-coefficient $s$. Then in Fig.~\ref{fig:fig1} (d), (e) and (f), we performed the extrapolation of $s$ by gradually removing the smallest system size $L_{\min}$ in the fitting process, and one sees that $s$ of there curves nicely yield the $s=0.020(1)$. We note that for all extrapolations, the
process is stopped once the two smallest values of $1/L_{\min}$ converge. The final converged slope is obtained by averaging the results from the
last two points. 

At $K=K_c$, i.e., at the Gaussian fixed point, it is interesting to see that the slope in Fig.~\ref{fig:fig1} (b) is distinctively different from those Ising QCPs. Here we put the analytic value of $s=0.02567$ by hand in the figure, and one sees the data follow the expectation perfectly. Moreover, in Fig.~\ref{fig:fig1} (g), we performed the same $1/L_{\min}$ extrapolation and found that converged $s=0.025(1)$ steadily land at the theoretical value for the Gaussian fixed point (red dashed line). We emphasize that without our bubble PQMC and the incremental SWAP implementation in this basis, and the SCEE cancellation of the area-law term in QMC sampling, the 1/1000 difference of the $s=0.020(1)$ for (2+1)d Ising QCP and the $s=0.025(1)$ for (2+1)d Gaussian QCP, were not possible with conventional computations.

For $K = 160,320 > K_c$, the system exhibits a first-order phase transition, characterized by the coexistence of PM and FM phases. Our numerical results demonstrate that the universal corner coefficient $s$ vanishes, as shown in Fig.~\ref{fig:fig1} (c) and its extrapolation in Fig.~\ref{fig:fig1} (h). This observation can be understood from the superposition of the contributions from the PM and FM phases, both of which are product states with gapped spectra and expected to have vanishing individual corner coefficients. Consequently, at the first-order transition point, the overall value is found to be $s = 0$.

The converged values of the universal corner logarithmic coefficient $s$ along the phase boundary from the Ising QCP to the Gaussian QCP and eventually to the first-order transition are summarized in Fig.~\ref{fig:fig1} (c). To place our results in the context of previous studies, we further provide a systematic comparison in Table~\ref{tab:tab1}. The table lists previously reported estimates known to us of $s$ for four $90^\circ$ corners at the $(2+1)$d Ising QCP, together with the method used, the reported value, and the corresponding citation. We also include the exact analytic free-field result at the Gaussian fixed point and our numerical estimate at the Gaussian QCP. This comparison shows that our Ising result is consistent with the most precise previous estimates, while the agreement between our Gaussian result and the analytic value provides an important benchmark for the accuracy of our method.
\begin{table}[h!]
\centering 
\caption{Comparison of previous estimates of the universal corner coeffi $s$ for four $90^\circ$ corners at the $(2+1)$d Ising and Gaussian QCPs. Our results are highlighted in bold.}
\begin{tabular}{@{}lcc@{}}
\toprule
Method & Universality class & $s$ \\
\midrule
Extended-ensemble estimator \cite{humeniukQuantum2012} 
& Ising & $0.003(1)$ \\
Series expansion \cite{singhThermodynamic2012} 
& Ising & $0.022(2)$ \\
Ratio-trick PQMC \cite{InglisEntanglement2013} 
& Ising & $0.024(8)$ \\
Numerical linked-cluster expansion \cite{kallinEntanglement2013} 
& Ising & $0.0212$ \\
\textbf{This work} 
& \textbf{Ising} & $\boldsymbol{0.020(1)}$ \\
\midrule
Analytic free-field calculation \cite{UniversalCasini2007} 
& Gaussian & $0.02567$ \\
\textbf{This work} 
& \textbf{Gaussian} & $\boldsymbol{0.025(1)}$ \\
\bottomrule
\end{tabular}
\label{tab:tab1}
\end{table}

\subsection{Error Analysis for the Universal Corner Coefficient $s$}
We now systematically describe the error analysis for the extracted universal corner coefficient $s$. Each value of $S_s$ defined as Eq.~\eqref{eq:SCEE definition} is computed independently via QMC sampling. Statistical errors on $S_s$ are estimated using binning analysis with block sizes chosen to eliminate autocorrelations in the Monte Carlo time series. These errors are propagated through a linear fit of $S_s$ versus $\ln L$ to extract the statistical error on $s$. The linear fit is performed using the \texttt{curve\_fit} function from the SciPy library \cite{scipy_curve_fit}, which accounts for the input errors on $S_s$. To obtain the thermodynamic-limit value of $s$, we perform a sequential extrapolation by gradually excluding the smallest system sizes from the fit (as shown in Fig.~\ref{fig:fig1} (d-h)). The final error on $s$ combines the statistical error from the fit with the uncertainty. We have performed extensive convergence tests to confirm that $m=8L^3$ is sufficient to eliminate all systematic error from the ground-state projection (see Appendix~\ref{appE} for details). The residual systematic error from $m$ is smaller than the statistical error for all system sizes and parameter values studied.

\section{Analytic results and Algorithmic advances of the Bubble PQMC}
\label{sec:method}
We now explain the analytic computation of the universal corner EE at the Gaussian fixed point (Sec.~\ref{sec:methodA}), and algorithmic advancements of {\it bubble basis} PQMC (Sec.~\ref{sec:methodB}) and the superior incremental SWAP EE computation in the bubble basis (Sec.~\ref {sec:methodC}), over the previous $\sigma^z$-basis.

\begin{figure}[htp!]
\includegraphics[width=\linewidth]{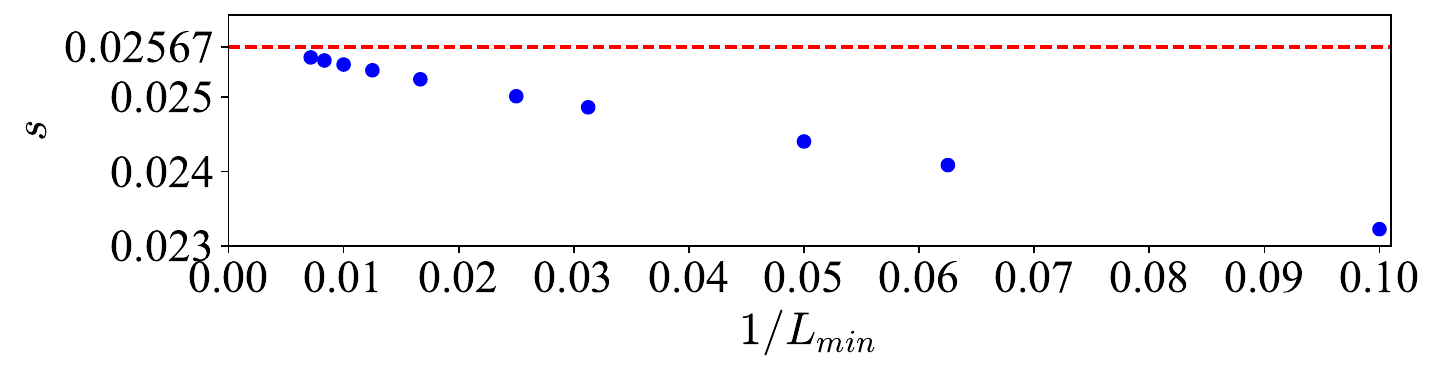}
\caption{\textbf{Universal corner term in a free theory.} In 2d square lattice with the entanglement area having four 90$^\circ$ corners, one can compute the universal term $s$ in Eq.~\eqref{eq:eq7} and extrapolate the value with the inverse linear size of the region $1/L_{\min}$, the extrapolated value is $s=0.02567$. The red dashed line is the reference value from literature~\cite{UniversalCasini2007,UniversalHelmes2016}.}
\label{fig:theory}
\end{figure}

\subsection{EE for free theory in a lattice}
\label{sec:methodA}
Following Refs.~\cite{peschelCalculation2003,UniversalCasini2007,UniversalHelmes2016}, here we perform a calculation of the universal corner log-term in EE for a free scalar theory on a square lattice. To this end, we evaluate the EE for a massless real scalar with the Hamiltonian
\begin{equation}
    H=\frac{1}{2}\sum_{i}\pi^2_i+\frac{1}{2}\sum_{\langle i,j\rangle} (\phi_i -\phi_j)^2=\frac{1}{2}\sum_{i}\pi^2_i+\frac{1}{2}\sum_{ i,j} \phi_i M_{i,j}\phi_j,
\end{equation}
where $\phi_i$ and $\pi_i$ are the free boson field and its canonical momentum, with the canonical commutation relation $[\phi_i,\pi_j]=i \delta_{i,j}$. The model is placed on a $L\times L$ lattice, and $M$ is the $N\times N$ (with $N=L^2$)  lattice Laplacian matrix.  In order to avoid zero mode, we impose anti-periodic boundary condition for $\phi$ in the $y$ direction (and periodic in $x$). From the Hamiltonian, we find the following correlators
\begin{eqnarray}
    X_{ij}&=&\langle \phi_i\phi_j\rangle=\frac{1}{2} (M^{-1/2})_{ij} \\ 
    P_{ij}&=&\langle \pi_i\pi_j \rangle =\frac{1}{2}(M^{1/2})_{ij}.
\end{eqnarray}

The entanglement entropy can be obtained using the following method. Denote the restriction of $X$ to a region $A$ by $X_A$ (and similarly define $P_A$). Diagonalize the matrix $\sqrt{X_AP_A}$, with eigenvalues $\epsilon_n$. 
 The 2nd R\'enyi EE is given by 
\begin{eqnarray}\label{eq:eq9}
    S^A_2=-\sum_n\ln\left((\epsilon_n+\frac{1}{2})^2-(\epsilon_n-\frac{1}{2})^2\right)=-\sum_n\ln(2\epsilon_n).
\end{eqnarray}

In order to cancel the leading perimeter term, we also employ the SCEE scheme and take the difference between them:
 \begin{equation}
S_s= S_2^{A_1}(L)-S_2^{A_2}(L)=s\ln L + \gamma+O(1/L).
     \label{eq:eq7}
 \end{equation}
 We gradually skip the smaller system sizes in the fitting, i.e., by throwing away linear size $L< L_{\min}$. The converged universal corner term $s=0.02567$ for four 90$^\circ$ corners, is obtained as shown in Fig.~\ref{fig:theory}. This is the reference value we use to compare the $s$ from QMC simulation at the (2+1)d Ising and Gaussian QCPs. In fact, at the Gaussian QCP at $K=K_c$ this value is indeed obtained from the PQMC simulaition, as shown in Fig.~\ref{fig:fig1} (b), (c) and (g). 

\subsection{Projector quantum Monte Carlo in Bubble basis}
\label{sec:methodB}
The PQMC method for bosonic/spin system was introduced by Sandvik in 2005 for SU(2) quantum spins using the valence-bond basis~\cite{sandvikGround2005}; later, a combined spin and valence-bond basis for SU(2) quantum spins was proposed to enable more efficient simulations~\cite{sandvikLoop2010}. In 2013, Inglis and Melko adapted PQMC to the TFIM in the $\sigma_z$ basis, enabling the calculation of the Rényi EE in such basis~\cite{InglisEntanglement2013}. 

In the ground state, the expectation value of an operator $\mathcal{O}$ is
\begin{equation}
\langle \mathcal{O} \rangle  = \frac{1}{Z} \langle \psi^0 | \mathcal{O} | \psi^0 \rangle,
\label{eq:eq12}
\end{equation}
where $| \psi^0 \rangle$ is the unnormalized ground state wavefunction and $Z =  \langle \psi^0 | \psi^0 \rangle$ is the normalization.
In PQMC framework, $| \psi^0 \rangle$ is estimated by a projection procedure on a {\it trial} state $|\alpha^0 \rangle$ as
\begin{equation}
(-H)^m |\alpha^0 \rangle \rightarrow | \psi^0 \rangle \hspace{2mm} {\rm , as} \hspace{2mm} m \rightarrow \infty. \nonumber
\end{equation}
To understand this projection procedure, one can write $|\alpha^0 \rangle$ as linear superposition of energy eigenstates of Hamiltonian $|\alpha^0 \rangle= \sum_i c_i | \psi^i \rangle$, then
\begin{equation}
    (-H)^m |\alpha^0 \rangle = (-E_0)^m \left[{  c_0 | \psi^0 \rangle + c_1 \left({ \frac{E_1}{E_0} }\right)^m| \psi^1 \rangle \cdots  }\right],
\end{equation}
where we assume that $E_0$ has the largest magnitude among all $E_i$ and this can be ensured by adding a sufficiently large negative constant to $H$.
Then, for large enough $m$, one can obtain $\langle \mathcal{O} \rangle$ as,
\begin{equation}
\langle \mathcal{O} \rangle  = \frac{\langle \alpha^0_l | (-H)^m  \mathcal{O}  (-H)^m | \alpha^0_r \rangle}{\langle \alpha^0_l | (-H)^m (-H)^m | \alpha^0_r  \rangle} ,
\label{eq:eq14}
\end{equation}
where $\langle \alpha^0_l |$ and $| \alpha^0_r  \rangle$ represent the left and right trial wavefunctions, they may be identical or distinct as there is no need for orthogonalisation.

To introduce the PQMC sampling, we rewrite the Hamiltonian (up to an additive constant) as,
\begin{equation}\label{eq:hamiltonain}
H=-J\sum_{N_J} H_{J}-h\sum_{N_h} H_{h}-K\sum_{N_K} H_{K},
\end{equation}
where $H_J=(\sigma^z_i \sigma^z_j+\mathbf{I})$ is the bond operator and $H_h=(\sigma^{+}_{i}+\sigma^{-}_{i}+\mathbf{I})$ is the site operator with 
$\sigma^{+} = (\sigma^x+i\sigma^y)/2$ and $\sigma^{-} = (\sigma^x-i\sigma^y)/2$, $H_K = (\sigma_a^z\sigma_d^z+\mathbf{I}) (\sigma_b^z\sigma_c^z+\mathbf{I})$ represents the four-body operator, and  $\mathbf{I}$ the identity operator. For a 2D system, $N_J=2N$ is the total number of bonds, $N_K=N$ and $N_h=N$.  In practice, we use $m=8L^3$ for all parameters, which is sufficient for capturing ground-state properties.

We sample the operators $H_{a}$, with $a\in \{J, h, K\}$ by evaluating 
\begin{equation}
\langle \mathcal{O} \rangle  = \frac{\sum_{\{c\}}  \langle \alpha^0_l | \prod_{i=1}^m H_{a_i}  \mathcal{O}  \prod_{j=1}^m H_{a_j} | \alpha^0_r \rangle}{\sum_{\{c\}}  \langle \alpha^0_l | \prod_{i=1}^m H_{a_i} \prod_{j=1}^m H_{a_j} | \alpha^0_r  \rangle} = \frac{\sum_{\{ c \}} \mathcal{W}_c O_c }{\sum_{\{ c \}} \mathcal{W}_c},
\label{eq:eq16}
\end{equation}
where $c\equiv {\{ a_1,a_2,\cdots, a_{2m} \}}$ is a configuration representing a set of operators in $2m$ projection space, the configuration weight is given as 
\begin{equation}\label{eq:weight}
    \mathcal{W}_c  =  \langle \alpha^0_l | \prod_{i=1}^{2m} H_{a_i}| \alpha^0_r  \rangle,
\end{equation}
and the estimator is 
\begin{equation}\label{eq:obs}
   O_c = { \langle \alpha^0_l | \prod_{i=1}^m H_{a_i}  \mathcal{O}  \prod_{j=1}^m H_{a_j} | \alpha^0_r \rangle}/\mathcal{W}_c. 
\end{equation}

In principle, the PQMC framework is basis-independent. In practice, however, the choice of basis is crucial for constructing a trial wave function that, when acted upon by the operator $H_a$, prevents branching of the update line and ensures the weights $\mathcal{W}_c$ remain non-negative—thereby circumventing the sign problem.  
As shown in Ref.~\cite{InglisEntanglement2013}, the $\sigma_z$ basis satisfies these requirements for conventional observables such as energy and correlation functions, leading to an algorithm whose computational complexity scales as $\mathcal{O}(m)$.  
However, precise calculation of the EE—which is known to be an exponential observable~\cite{zhangIntegral2024}—often requires specialized techniques such as the incremental approach to overcome severe statistical fluctuations in direct sampling~\cite{zhouIncremental2024,zhangIntegral2024, liaoUniversal2025}. Applying the incremental approach requires incorporating the sampled value of the SWAP operator into the update weights. However, within the conventional $\sigma_z$ basis PQMC framework, the sampled value of the SWAP operator can only be evaluated in the middle slice of the projection space. Consequently, implementing the incremental technique for computing the EE of the TFIM in this basis leads to a substantial increase in computational cost. As we analyze below, the overall scaling rises unfavorably to $\mathcal{O}(m^2)$, which severely limits the system sizes for which reliable EE calculations are feasible.

To enable accurate EE calculation for considerably larger systems, thereby allowing more robust finite-size scaling and credible revelation of universal properties, we introduce a new PQMC framework formulated in a {\it bubble basis}. 
As shown below, when combined with the incremental technique, our bubble basis PQMC substantially reduces the computational overhead to $\mathcal{O}(mP)$ where $P$ scales as a power-law in the linear system size $L$ in the FM phase and at the critical point, but remains $\mathcal{O}(1)$ in the PM phase, as shown in Fig.~\ref{fig:fig5}. This improvement arises because the sampled value of the SWAP operator can be evaluated in any projector slice. This advancement dramatically enhances the efficiency of precise EE computations. Below we explain the concept of the bubble basis.

\begin{figure}[htp!]
\includegraphics[width=0.8\linewidth]{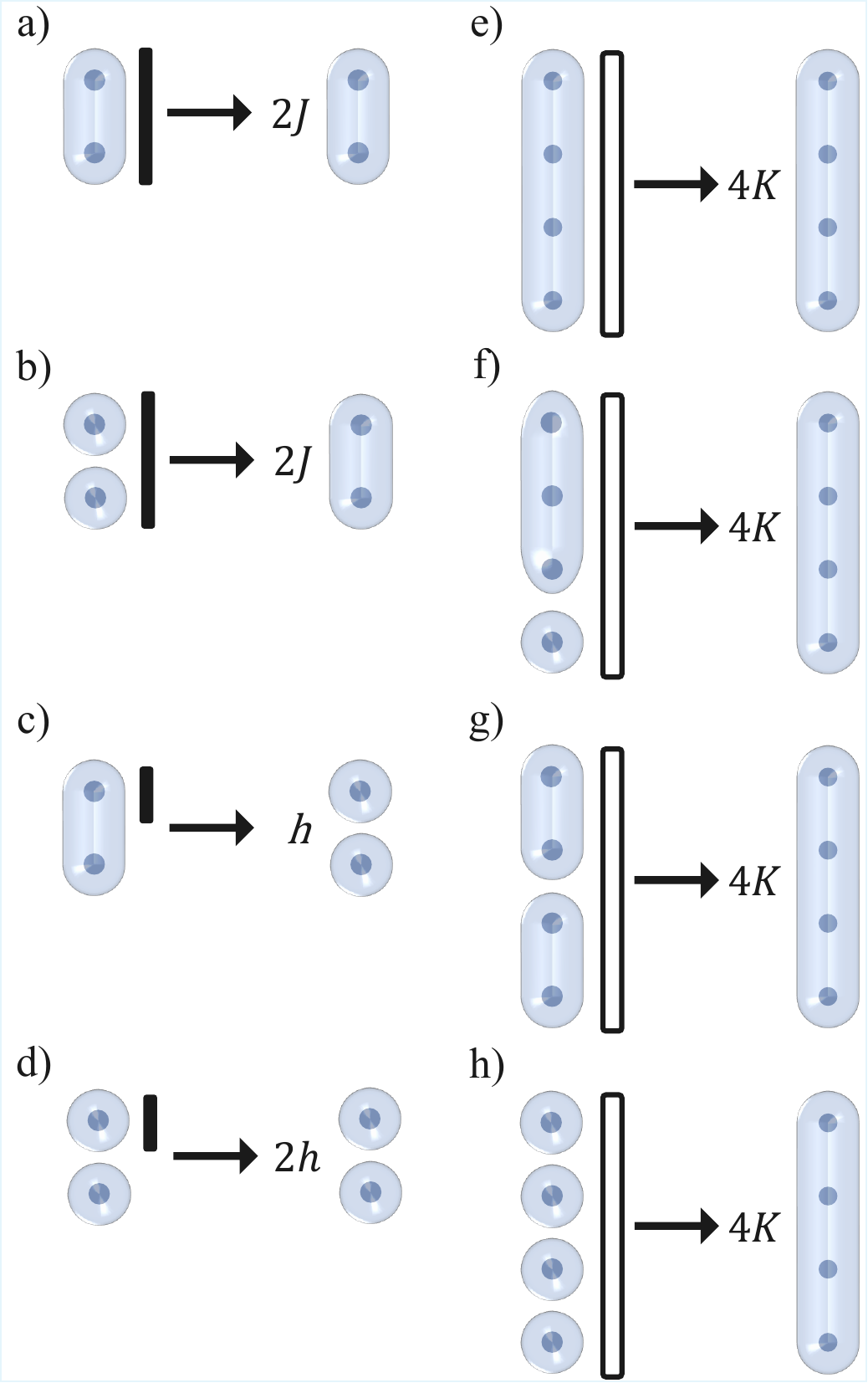}
\caption{\textbf{Illustration of the bubble basis and its operator evolution. } a) and b) show the action of $H_J$, c) and d) show the action of $H_h$ and e) - h) show the action of $H_K$ on the bubble basis. The $H_K$ acts on a square of four spins, but for illustration purposes, it is compressed and shown as a one-dimensional bar here. After acting the operators on the basis, they would give different constants as described in Eq. \eqref{eq:weight} and these constants are also shown next to the bubbles.}
\label{fig:fig2}
\end{figure}

\begin{figure}[htp!]
    \includegraphics[width=\columnwidth]{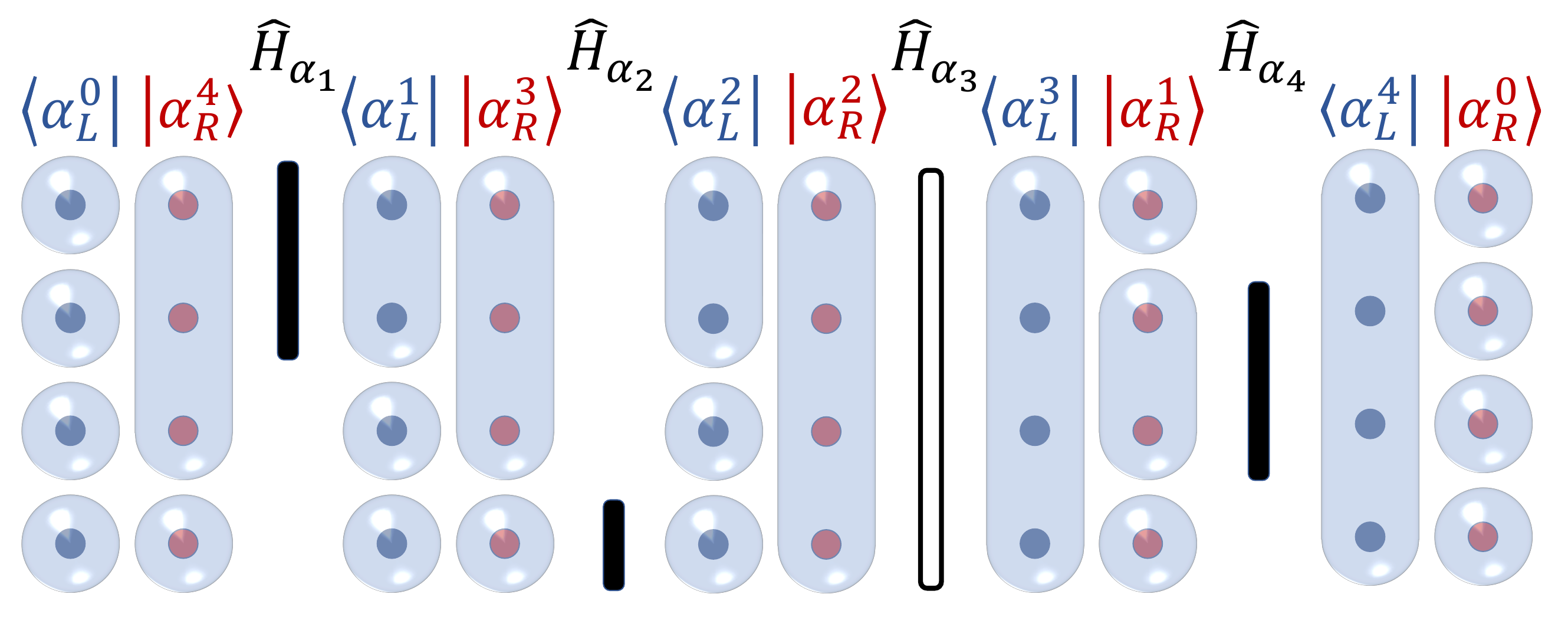}
    \caption{\textbf{Propagation of the operators. }This graph gives a pictorial representation of Eq. \eqref{eq:timeslice}. It shows a configuration with $m=2$ and the operators $H_{a_1}$ - $H_{a_4}$ are $H_J$, $H_h$, $H_k$ and $H_j$ respectively. The actions of different operators on the bubble basis are illustrated in Fig. \ref{fig:fig1}, and the constants are neglected here. }
    \label{fig:fig3}
\end{figure}

As illustrated in Fig.~\ref{fig:fig2}, a state $| \alpha^0  \rangle$ in bubble basis is employed as the trial wave function. The term ``bubble'' is used because, when visualizing the evolution of the state, the loops formed by different lattice sites resemble bubbles floating in the configuration space (see Appendix~\ref{appD} for more graphic interpretation). Within a single bubble, all sites share the same spin orientation, either all up or all down.
In other words, a bubble state is a superposition of $\sigma^z$ product states. 
For example, for the right trail wave function in Fig.~\ref{fig:fig3}, in a four-sites system with four bubbles, we have
$| \alpha^0_R  \rangle = | ( \uparrow  + \downarrow  ) \rangle  ( \uparrow  + \downarrow  ) \rangle  ( \uparrow  + \downarrow  ) \rangle  ( \uparrow  + \downarrow  ) \rangle$. 
It is straightforward to verify that the application of a bond operator $H_J$, a site operator $H_h$, or a four-body operator $H_K$ to a bubble state results in another bubble state, with no branching occurring, as shown in panels (a)-(h) in Fig.~\ref{fig:fig2}.

\begin{figure}[htp!]
	\centering
	\includegraphics[width=0.16\linewidth]{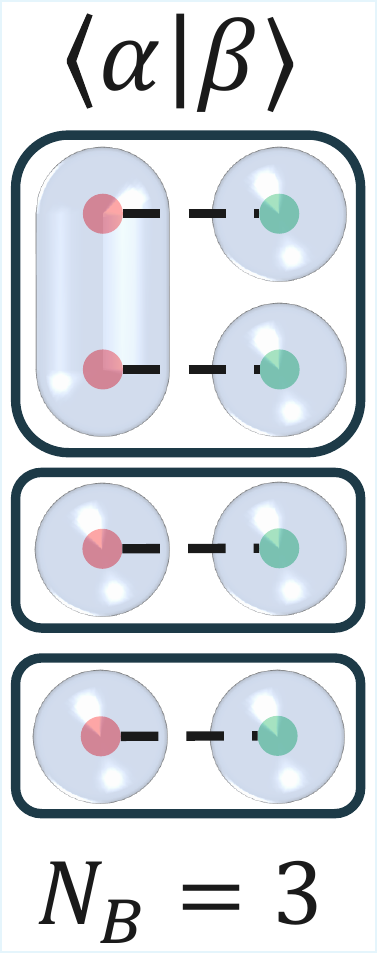}
	\caption{\textbf{Illustration of the overlap of bubble basis.} Here $\langle \alpha | = \langle  (\uparrow \uparrow + \downarrow \downarrow) (\uparrow + \downarrow) (\uparrow + \downarrow) |$ and $| \beta \rangle = | (\uparrow + \downarrow) (\uparrow + \downarrow) (\uparrow + \downarrow) (\uparrow + \downarrow) \rangle$. The dash lines represent the one-to-one site corresponding notes. In this example, the black solid lines are the link-bubbles and the number of independent link-bubbles is $N_B = 3$. }
	\label{fig:fig4}
\end{figure}

\begin{figure}[htp!]
	\centering
	\includegraphics[width=0.9\linewidth]{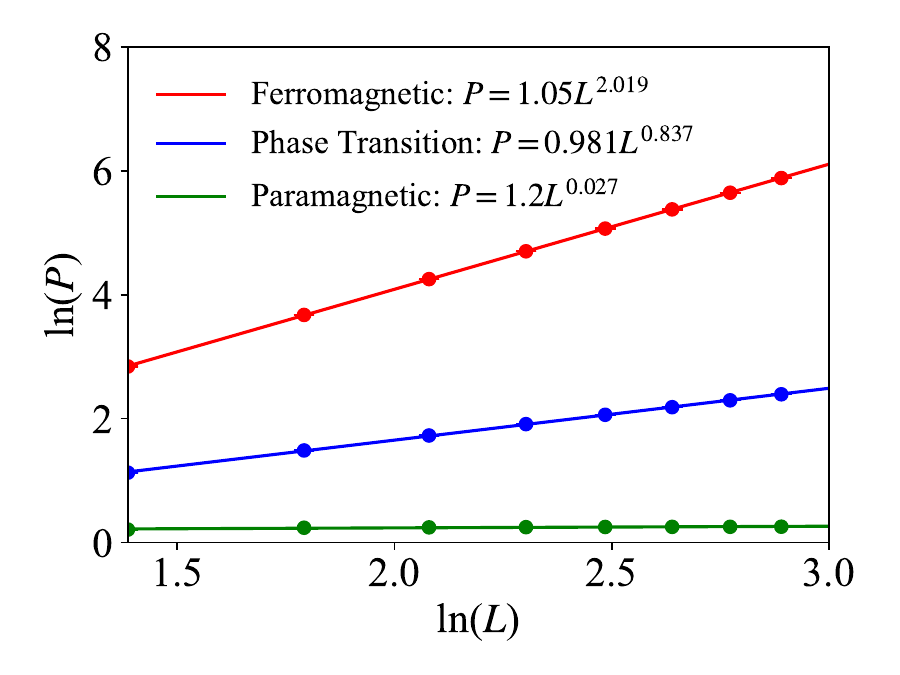}
    \caption{\textbf{Scaling behavior of the bubble size parameter $P$ across different phases and quantum critical points at $K=0$ (a) and $K=16.02$ (b).} Inside the FM phase at $K=0$, $P \sim L^{2.019}$; at the Ising QCP ($K=0$), $P \sim L^{0.837}$; inside the PM phase at $K=0$, $P\sim L^{0.027}$ (almost a constant). At the Gaussian QCP ($K=16.02$), we observe $P \sim L^{0.992}$. In all cases, $P \ll m = 8L^3$ in the thermodynamic limit.}
	\label{fig:fig5}
\end{figure}

It is straightforward to verify that the overlap of two bubble states is 
\begin{equation}
    \langle \alpha \vert \beta\rangle = 2^{N_{B}},
\end{equation}
where $N_B$ is the number of link-bubbles formed by connecting the bubbles between $ \langle \alpha \vert $ and $\vert \beta\rangle$ via their one-to-one site correspoding nodes and an example is shown in Fig.~\ref{fig:fig4}.

To estimate the observable $\langle \alpha |  \mathcal{O} | \beta \rangle$, one approach is to first apply the operator $\mathcal{O}$ to the state $|\beta\rangle$, evolving it to a new bubble state $|\gamma\rangle$, and then compute the overlap $\langle \alpha | \gamma \rangle$.  
Alternatively, one can work in the $\sigma^z$ basis: since bubble states are linear superpositions in the $\sigma^z$ basis, the expectation value $\langle \alpha |  \mathcal{O} | \beta \rangle$ can also be expressed as a weighted sum of expectation values of $\mathcal{O}$ in the $\sigma^z$ basis (detailed explanation is given in Appendix~\ref{appC}).

To update a configuration $c$ to $c^\prime$ according to the Metropolis algorithm within PQMC, calculating the configuration weight ratio $\mathcal{W}_{c^\prime}/\mathcal{W}_c$ is a key step. This ratio determines the acceptance probability for the proposed move.
A direct calculation of the individual weights $\mathcal{W}_c$ and $\mathcal{W}_{c^\prime}$ using the explicit Eq.~\eqref{eq:weight} typically involves computations for projection evolution and the state overlaps. 
For a local update of the configuration--where only a single operator $H_{a_j}$ is changed to $H_{a_j}^\prime$ with select rate $\frac{{a_j}^\prime N_{{a_j}^\prime}}{(JN_J+hN_h+KN_K)}$ at the $j$-th projection time slice, leaving all other operators unchanged--this direct approach can lead to a computational complexity of $\mathcal{O}(m+N)$. 
Here $m$ and $N$ come from projection evolution and the state overlaps, separately.
To achieve a more efficient calculation, the recently developed {\it propagating update strategy} by us~\cite{zhouIncremental2024} is used.

The key idea of the propagating update strategy is that, for a local update, Eq.~\eqref{eq:weight} can be reformulated as, 
\begin{equation}\label{eq:timeslice}
\left\langle \alpha_{l}^0\left|\prod\limits_{i=1}^{2m} H_{a_i} \right| \alpha_r^0 \right\rangle = 2^{\left( n_l^{j-1}+n_r^{2m-j} \right)} \left\langle \alpha_{l}^{j-1}\vert H_{a_j} \vert \alpha_r^{2m-j} \right\rangle,
\end{equation}
where the value of $n_l^{j-1}$ is determined by the sequence of operator actions encountered during the projector process $\langle \alpha_l^{j-1} \vert$, which evolves from $\langle \alpha_l^0\vert  \prod_{i=1}^{j-1} H_{a_j}$, as illustrated in Fig.~\ref{fig:fig3}. 
According to Fig.~\ref{fig:fig2}, each operator action corresponding to panels (a), (b), and (d) increments $n_l^{j-1}$ by 1. Actions represented by panel (c) contribute 0, while those shown in panels (e) to (h) increment $n_l^{j-1}$ by 2.
Conversely, $n_r^{2m-j}$ represents an analogous count for the process $\vert \alpha_r^{2m-j} \rangle$, which evolves from $\prod_{i=j+1}^{2m} H_{a_j} \vert \alpha_l^0\rangle$.

For a local update from $H_{a_j}$ to $H_{a_j}^\prime$, the weight ratio can be calculated straightforwardly as 
\begin{equation}\label{eq:ratio}
\frac{\mathcal{W}_{c^\prime}}{\mathcal{W}_{c}} = \frac{\left\langle \alpha_{l}^{j-1}\vert H_{a_j}^\prime \vert \alpha_r^{2m-j} \right\rangle}{\left\langle \alpha_{l}^{j-1}\vert H_{a_j} \vert \alpha_r^{2m-j} \right\rangle}.
\end{equation}
Since this approach avoids recomputing the entire propagation, the computational complexity for a local update is reduced from $\mathcal{O}(m+N)$ to $\mathcal{O}(N)$, where $\mathcal{O}(N)$ arises from the overlap calculation. 
The evaluation of the ratio can be further accelerated by introducing an auxiliary state, leading to the reformulated expression,
\begin{equation}\label{eq:ratiofast}
\frac{\mathcal{W}_{c^\prime}}{\mathcal{W}_{c}} = \frac{\left\langle \alpha_{l}^{j-1}\vert H_{a_j}^\prime \vert \alpha_r^{2m-j} \right\rangle / \left\langle \alpha_{l}^{j-1}\vert \alpha_r^{2m-j} \right\rangle }{\left\langle \alpha_{l}^{j-1}\vert H_{a_j} \vert \alpha_r^{2m-j} \right\rangle / \left\langle \alpha_{l}^{j-1}\vert \alpha_r^{2m-j} \right\rangle  }.
\end{equation}
As illustrated in Fig.~\ref{fig:fig2}, the action of a bond operator $H_J$ either merges two bubbles into one or leaves a bubble unchanged. If we express the overlap as $\left\langle \alpha_{l}^{j-1}\vert H_{a_j} \vert \alpha_r^{2m-j} \right\rangle = 2^{N_B^\prime}$ and $ \left\langle \alpha_{l}^{j-1}\vert \alpha_r^{2m-j} \right\rangle = 2^{N_B} $, then the difference $N_B^\prime-N_B$ is either $-1$ or $0$.
Consequently, $\left\langle \alpha_{l}^{j-1}\vert H_{a_j} \vert \alpha_r^{2m-j} \right\rangle / \left\langle \alpha_{l}^{j-1}\vert \alpha_r^{2m-j} \right\rangle $ takes a value of $2^{-1}$ or $2^{0}$  when $H_{a_j}$ represents a bond operator.
Similarly, a site operator $H_h$ either splits one bubble into two or leaves it unchanged, resulting in possible values of $2^{1}$ or $2^{0}$ for $\left\langle \alpha_{l}^{j-1}\vert H_{a_j} \vert \alpha_r^{2m-j} \right\rangle / \left\langle \alpha_{l}^{j-1}\vert \alpha_r^{2m-j} \right\rangle $ when  $H_{a_j}$ represents a site operator. 
A four-body $K$-term operator can be treated as the product of two bond operators.
Furthermore, determining the value of $N_B^\prime-N_B$ only requires consideration of the link-bubbles that contain the sites involved in the operator.
Therefore, the computational complexity of the weight ratio calculation via the accelerated Eq.~\eqref{eq:ratiofast} can be expressed as $O(P)$, where $P$ denotes the statistical average of the number of lattice sites within the link-bubbles connected to the operator being updated. 

In our implementation, we store the states of the left and right bubbles. After updating the configuration at the $j$‑th projector slice, the left bubble state $\langle \alpha_l^{j-1} |$ and the right bubble state $| \alpha_r^{2m-j} \rangle$ propagate to $\langle \alpha_l^{j} |$and $| \alpha_r^{2m-j-1} \rangle$, respectively. The operator at the $(j+1)$-th projector slice is then updated. This process is carried out sequentially from the first slice to the $2m$‑th slice, and then back from the $2m$‑th slice to the first slice. We repeat this back‑and‑forth propagation, which is why the method is called a propagating update. Further implementation details are provided in Appendix~\ref{appD}.

As illustrated in Fig.~\ref{fig:fig5}, $P$ exhibits a power-law scaling with the system size $L$ in the FM phase and at both the Ising and Gaussian QCPs, while it remains $O(1)$ in the PM phase.
In the FM phase at $K=0$, the tendency for all lattice sites to form a single collective ``bubble'' implies that the weight ratio calculation must account for the entire lattice. This is consistent with the observed scaling $P \sim L^2$, which equals the total number of lattice sites $N$. Conversely, in the PM phase at $K=0$, individual sites or small local clusters tend to form independent bubbles. Consequently, the value of $P$ becomes a constant independent of $L$. Notably, at the Ising QCP ($K=0$), numerical results indicate that $P$ follows a power-law scaling with $L$ with an exponent of approximately $0.837$. At the Gaussian QCP ($K=16.02$), we observe a power-law scaling $P \sim L^{0.992}$. While this exponent is larger than that at the Ising QCP, $P$ remains vastly smaller than the projection length $m=8L^3$ in both cases: the ratio $P/m \sim L^{-2.163}$ at the Ising QCP and $P/m \sim L^{-2.008}$ at the Gaussian QCP, both of which vanish rapidly in the thermodynamic limit. This leads to a significant computational speedup compared to the $\sigma^z$-basis approach. When considering the full projection time slice, the total computational complexity of the PQMC method with the bubble basis is therefore $O(mP)$, where $m$ is the number of time slices.

The power-law scaling of $P$ with system size at quantum critical points is a universal, model-independent property. In contrast, the specific numerical value of the scaling exponent varies across different universality classes and may be model-dependent. Regarding whether the specific numerical value is universal for a given universality class, we have no numerical results to confirm this at present, and we note that this remains an interesting open question.

\begin{figure}[htp!]
	\centering
	\includegraphics[width=0.95\columnwidth]{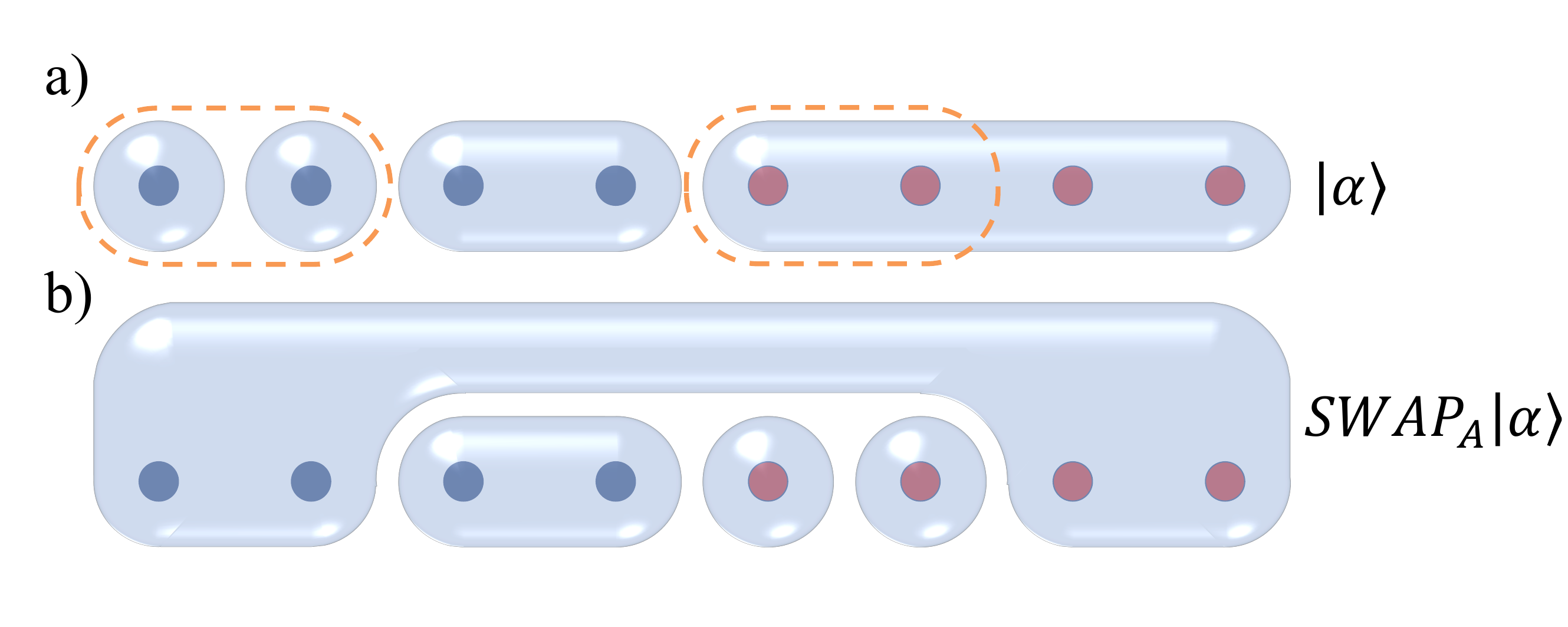}
	\caption{\textbf{SWAP operator actions on bubble basis. } An eight-site chain with two interacting copies (blue and red color sites) before (a) and after (b) $SWAP_A$. The orange dotted line circles the entangled region $A$. The bubbles in the graph follow the same definition as in the main text.}
	\label{fig:fig6} 
\end{figure}

We have successfully applied the developed PQMC algorithm in the bubble basis to compute the phase diagram of the transverse field Ising model, as shown in Appendix~\ref{appC}. Our results indicate that a continuous phase transition from a FM to a PM state occurs at a critical field of $h_c = 3.044(2)$. The critical exponents obtained numerically demonstrate that this phase transition belongs to the (2+1)d Ising universality class, which is consistent with previous literature~\cite{pfeutyIsing1971,hesselmannThermal2016,xuNoFermi2017}. 

\subsection{EE computation in the Bubble PQMC}
\label{sec:methodC}
As discussed above, the total computational complexity of PQMC with the bubble basis is $\mathcal{O}(mP)$. In the PM phase, $P \sim 1$, whereas in the FM phase and at QMC, $P \sim L^{\alpha}$. In our case, $\alpha\simeq 2$ in the FM phase and $\alpha\simeq 0.8$ at QCP (cf. Fig.~\ref{fig:fig5}). The PQMC with the $\sigma_z$ basis developed in Ref.~\cite{InglisEntanglement2013} has computational complexity $O(m)$ across all phases for general measurements such as energy, magnetization, and the Binder cumulant. Consequently, our method in bubble basis is as efficient as the $\sigma_z$ basis PQMC in the PM phase, but it is less efficient in the FM phase and at the QCP for general measurements.
This explains why we used the $\sigma_z$ to obtain the raw data presented in Fig.~\ref{fig:fig8}. 
However, for the computing of EE, the bubble basis provides a significant efficiency gain. It reduces the complexity from $\mathcal{O}(m^2)$ for EE computation in $\sigma_z$ basis to $\mathcal{O}(mP)$ across all phases, as detailed in the following explanation.

The EE can be estimated by the expectation value of $\widehat{\text{SWAP}}_A$ operator using a replica trick in QMC~\cite{CalabreseEntanglement2004,hastingsMeasuring2010,InglisEntanglement2013}, 
\begin{equation}
    e^{-S^A_2} = \frac{ \sum_{\{c_1\},\{c_2\}} \mathcal{W}_{12}  \ ( \text{SWAP}_A / \mathcal{W}_{12}   ) }{\sum_{\{c_1\},\{c_2\}}  \mathcal{W}_{12} } 
    \label{eq:eq22}
\end{equation}
where $ \mathcal{W}_{12} =  \mathcal{W}_{c_1} \mathcal{W}_{c_2}$ is the weight product of two replicas, $A$ is the entanglement region as shown in the inset of Fig.~\ref{fig:fig1} (a) and 
\begin{equation}
    \text{SWAP}_A =  \langle \alpha^0_{l,12} | \prod_{i=1}^m H_{a_i,1} H_{a_i,2} \widehat{\text{SWAP}}_A  \prod_{j=1}^m H_{a_j,1} H_{a_j,2}| \alpha^0_{r,12} \rangle ,
\end{equation}
where $\langle \alpha^0_{l,12} |= \langle \alpha^0_{l,1} | \otimes \langle \alpha^0_{l,2} |$ 
and 
$| \alpha^0_{r,12} \rangle = | \alpha^0_{r,1} \rangle \otimes | \alpha^0_{r,2} \rangle$
are the left and right trail wavefunctions of a system containing two replicas, respectively.

Recent studies have established that EE is an exponential observable \cite{zhouIncremental2024,zhangIntegral2024, liaoUniversal2025}, and directly evaluating $\text{SWAP}_A$ is plagued by convergence problems. More precisely, in a naive computation scheme of the ratio of replica partition functions~\cite{InglisEntanglement2013,stoudenmireCorner2014,CornerKallin2014,EntanglementHelmes2014}, the coefficient of variation of the obtained EE will increase exponentially with the system sizes~\cite{panStable2023,liaoUniversal2025}, rendering the computation complexity grows exponentially if one wants to control the relative percentage of standard deviation over the mean value. To overcome these issues, the power incremental algorithm provides an elegant and straightforward operational approach~\cite{zhouIncremental2024, liaoUniversal2025}. 
Rather than estimating $\text{SWAP}_A$ directly, one should compute it in an incremental form 
\begin{equation}\label{eq:incremental}
 	 e^{-S_2^A}=\frac{Z(1)}{Z(0)}\frac{Z(2)}{Z(1)}\cdots\frac{Z({k+1})}{Z(k)} \cdots\frac{Z({n})}{Z({n} - 1)},
\end{equation}
where $Z\left(k\right)= \sum_{\{c_1\},\{c_2\}} \mathcal{W}_{12} ( \text{SWAP}_A/\mathcal{W}_{12}  ) ^{k/n}$. Here integer $k$ indexes the $k$-th increment out of a total of $n$ increments. The value of $n$ can be quantitatively determined based on the scaling of $\log( \text{SWAP}_A)$, which scales with the linear system size $L$ in a power-law.
With $Z(0) = \sum_{\{c_1\},\{c_2\}} \mathcal{W}_{12} $ and  $Z({n}) = \sum_{\{c_1\},\{c_2\}} \mathcal{W}_{12} ( \text{SWAP}_A / \mathcal{W}_{12}  )$, the final product naturally yields the definition of the 2nd R\'enyi EE.
Each incremental ratio can then be evaluated in parallel with PQMC as follows
\begin{equation}\label{eq:incre-slice}
\frac{Z\left({k+1}\right)}{Z\left(k\right)}=\frac{ \sum_{\{c_1\},\{c_2\}} \mathcal{W}_{12}^{1-k/n}  ( \text{SWAP}_A )^{k/ n} ( \text{SWAP}_A/\mathcal{W}_{12})^{1/n} }{ \sum_{\{c_1\},\{c_2\}} \mathcal{W}_{12}^{1-k/n} ( \text{SWAP}_A )^{k/n}}.
\end{equation}

Implementing this incremental approach, however, requires incorporating the sampled value of the SWAP operator into the update weights. Within the conventional $\sigma_z$-basis PQMC framework, the value of SWAP$_A$ can only be determined at the central projection time slice. The reason is that evaluating SWAP$_A$ first requires building clusters for off-diagonal updates; its value is then determined by counting how many independent clusters cross the central time slice, as described in Ref.~\cite{InglisEntanglement2013}.
Consequently, for every local update at any projection time slice, the configuration must be propagated to the central time slice to reconstruct the clusters and re‑evaluate the crossing count. This procedure must be repeated for each of the $O(m)$ local updates, and each propagation step itself scales as $O(m)$, leading to an overall computational cost of $O(m^2)$. This quadratic scaling is prohibitively expensive.

In contrast, the bubble basis PQMC method provides a key advantage. The action of the $\widehat{\text{SWAP}}$ operator on a bubble state produces another well-defined bubble state, as shown in Fig.~\ref{fig:fig6}. This property makes the propagating update strategy based on Eq.~\eqref{eq:ratiofast} straightforwardly applicable to the $\widehat{\text{SWAP}}$ operator during updates. The only additional overhead is performing a single update involving the $\widehat{\text{SWAP}}$ operator at the central time slice, whose cost is negligible compared to all other updates. Consequently, the total computational complexity for evaluating each ratio $\frac{Z(k+1)}{Z(k)}$ remains $\mathcal{O}(mP)$, making the bubble-basis PQMC approach significantly more efficient than the $\sigma^z$ basis method.

\section{Discussion}
\label{sec:Discussion}
Based on the description in Secs.~\ref{sec:results} and \ref{sec:method}, as for the three questions that were asked in the introduction: 
\begin{itemize}
    \item There is a lack of efficient algorithms for computing EE at (2+1)d using the ratio of replica partition functions, which can overcome the exponential growth of the coefficient of variation and the computational complexity.
    \item It is difficult to separate and access the intensive and subleading universal term of the corner log-coefficient from the extensive area law piece of the computed EE, as a function of increasing system sizes.
    \item There exists no previous attempt that within one QMC computation, the analytically known results of these universal terms can be recovered consistently and then extend the computation to the strongly coupling regime, at (2+1)d.
\end{itemize}
We believe these three questions are solved satisfactorily in this work. 

By designing a lattice model that hosts QCPs from (2+1)d Ising to Gaussian fixed point -- tricritical Ising at its upper critical dimension, and by efficiently computing the EE with the incremental SWAP PQMC algorithm in the bubble basis such that the leading area-law contributions are eliminated within one simulation and the sub-leading logarithmic term is promoted to the leading term. We obtained the precise quantitative values of the universal corner log-coefficient in EE for the (2+1)d Ising and Gaussian QCPs. We find that these two universalities are distinguishably different in their universal term of EE, through our high precision measurements, of 0.020(1) for the former and 0.025(1) for the latter for four 90$^\circ$ corners of the entanglement area. The exact matching with the analytic results~\cite{UniversalCasini2007,UniversalHelmes2016} at the Gaussian QCP, implies that our Ising value is by far the most accurate one among all the previous attempts.

The significance of this work can be appreciated in a broader perspective. Our computation scheme gives the future QMC and numerical computation of the EE at (2+1)d interacting systems a solid foundation, and the further methodology developments, either in the projector QMC or path-integral QMC and tensor-network or neural quantum state wavefunctions, can now use the present results and computational complexity as the necessary benchmark to evaluate their correctness and performance. 

Looking ahead, it is now the time to extend such a comparison at (2+1)d between the analytic results at Gaussian fixed point and the non-perturbative numerical results from large-scale computation to other entanglement measurements, such as multipartite entanglement~\cite{wangEntanglement2025,lyuMutiparty2025,songEntanglement2025}, quantum Fisher information~\cite{scheieWitnessing2021,zhouQuantum2025,shimokawaExperiment2025} and even quantum conditional mutual information for mixed state~\cite{wangAnalog2025}. With the hope that in (2+1) dimensions or higher, these entanglement properties can bridge the fundamental and quintessential properties of quantum states of matter from precise analytic calculation for often-time weakly interacting systems, to practical model computations for strongly correlated systems and eventually applications to realistic experiments for quantum materials. In this way, exotic quantum critical points and emerging quantum many-body phases, and their associated
theory, computation, and experimental detection, could be
expected to unite from the entanglement perspective.\\

\section*{Acknowledgements}
We thank Cenke Xu  for inspiring discussions on understanding the EE computation results over the years.  
YDL acknowledges support from National Natural Science Foundation of China (Grant No. 12404282), YDL and ZYM acknowledge support from General Program of the Guangdong Natural Science Foundation (Grant No. 2025A1515010337).
We acknowledge the support from the Research Grants Council (RGC) of Hong Kong (Project Nos. 17309822, C7037-22GF, 17302223, 17301924, 17301725), the ANR/RGC Joint Research Scheme sponsored by RGC of Hong Kong and French National Research Agency (Project No. A\_HKU703/22). 
We thank HPC2021 system under the Information Technology Services at the University of Hong Kong~\cite{hpc2021}, as well as the Beijing Paratera Tech Corp., Ltd~\cite{paratera} for providing HPC resources that have contributed to the research results reported within this paper. 
The authors gratefully acknowledge the computing time made available to them on the high-performance computer Barnard at the NHR Center of TU Dresden. This center is jointly supported by the Federal Ministry of Education and Research and the state governments participating in the National High-Performance Computing (NHR) joint funding program~\cite{dresdenurl}.

\appendix

\section*{appendix}

In Appendix~\ref{appA}, we provide the details in the RG flow analysis from the Gaussian fixed point to the Ising CFT. 

In Appendix~\ref{appB}, we present details of the non-equilibrium approach within the finite-temperature stochastic series expansion QMC framework—another method capable of extracting the subtracted SCEE. However, a key parameter in this approach, the quench step, could not previously be determined quantitatively, making it difficult to control the accuracy of the results. Here, we describe how to determine the quench step quantitatively, thereby placing the non-equilibrium method on a more complete footing. Furthermore, a direct comparison shows that the SCEE obtained from this refined non-equilibrium method agrees with results from the incremental SWAP algorithm.

In Appendix~\ref{appC}, we present an efficient technique for measuring observables using bubble basis PQMC and demonstrate the validity of the method via a critical analysis of the transverse field Ising model. 

In Appendix~\ref{appD}, we detail the implementation of the incremental SWAP technique for bubble basis PQMC, which yields a computational complexity of  $\mathcal{O}(mP)$.

\section{RG flow from the Gaussian fixed point to the Ising CFT} 
\label{appA}
In this section, we review the standard picture for the RG flow from the Gaussian fixed point to the Ising CFT. 
We first discuss Landau's mean field theory analysis~\cite{landau1937theory}. Since we are dealing with spontaneous $Z_2$ symmetry breaking, we can approximate the free energy as 
\begin{align}
    F(\phi)=\frac{m^2}{2}\phi^2+ \frac{\lambda_4}{4!} \phi^4+ \frac{\lambda_6}{6!} \phi^6+\cdots.
\end{align}
Here $\phi$ is the order parameter for the symmetry breaking. Assume $\lambda_6>0$. When $m^2>0$, the free energy has a single minimum, corresponding to the disordered phase (PM phase). 
When $m^2<0$, the free energy have two minima, corresponding to the $Z_2$ symmetry breaking phase (FM phase). 
The order of the phase transition depends on the sign of $\lambda_4$.
When  $\lambda_4>0$, the phase transition is second order. When $\lambda_4<0$, the phase transition is first order.
This phase diagram from Landau's theory agrees with the phase diagram in Fig.~\ref{fig:fig1} (a) from our numerical simulation. Changing $m^2$ and $\lambda_4$ corresponds to moving perpendicularly or parallel to the phase separation line respectively. 

Laudau's mean field theory analysis does not take into account the thermal/quantum fluctuations. To take them into account we should instead use the language of quantum field theory.
This was first understood in the seminal paper by Wilson and Fisher~\cite{PhysRevLett.28.240}, where they studied the renormalization group flow of the field theory: 
\begin{align}
    S=\int d^Dx\frac{1}{2}(\nabla_i\phi)^2+\frac{1}{2}m^2 \phi^2+\frac{1}{4!}\lambda_4\phi^4.
\end{align}
in $D=4-\epsilon$ dimensions. Setting $m^2=0$, the beta function of the coupling constant $\lambda_4$ is 
\begin{align}
    l\frac{d\lambda_4}{dl}=\epsilon \lambda_4-\frac{3}{16\pi^2} \lambda_4^2+\cdots.
\end{align}
Here $l$ is the characteristic length scale. 
The beta function tells us how the coupling constant evolves as we change the length scale.
The thermodynamic limit of lattice models corresponds to large scale physics, that is $l\rightarrow\infty$.
Higher order coupling terms, such as $\phi^6$ terms can be neglected because they are irrelevant, and becomes less important in the $l\rightarrow\infty$. The beta function have two fixed points
\begin{align}
    \lambda_4=0,\quad {\rm and}\quad \lambda_4 =\frac{16\pi^2}{3}\epsilon.
\end{align}
The fixed point at $\lambda_4=0$ is a Gaussian fixed point, which corresponds to the tricritical point in Fig.~\ref{fig:fig1} (a). 
The second fixed point is genuinely interacting, which corresponds to 2nd order Ising phase transition line in Fig.~\ref{fig:fig1} (a).
To study the physics in the physical dimension, we need to set $\epsilon=1$. 
At the Gaussian point, the subleading logarithmic term in the entanglement entropy arising
from sharp corners is well-established~\cite{UniversalCasini2007,UniversalHelmes2016}, due to the fact that Gaussian theories are easier to solve.
At the interacting point, on the other hand, we are not aware of any reliable theoretical calculation to get the logarithmic correction.
We benchmark our algorithm at the Gaussian fixed point, and then measure this quantity at the Ising fixed point.

\section{Non-equilibrium method for EE}  
\label{appB}

\subsection{Details of extracting SCEE with non-equilibrium method}
\label{appB1}
Here, we demonstrate the details of the non-equilibrium method. The discussion of this section follows closely with those in Refs~\cite{zhaoMeasuring2022, DEmidioEntanglement2020,liaoExtracting2024,songExtracting2024,songEvolution2025}.

We introduce a partition function definition $Z_{A_1 - A_2}(\lambda)$ parameterized by $\lambda$, where $Z(\lambda)$ is defined as the summation of collections for partition function $Z_B$ weighted by binomial factor 
\begin{equation}
    g_A(\lambda, N_B) = \lambda^{N_B} (1 - \lambda)^{N_{A_1} - N_{A_2} - N_B},
\end{equation}
where in the context of lattice model, $N_{A_1}$ is the number of sites in entanglement region $A_1$ and $N_B$ is the number of sites in the growing region. Typically, $N_B \le N_{A_1} - N_{A_2}$. The definitions of $A_1$ and $A_2$ are illustrated in Fig.~\ref{fig:fig1} (a).

The partition function can therefore be written as, 
\begin{equation}
\begin{aligned}
    Z_{A_1 - A_2}(\lambda) &= \sum_{B  \subseteq (A_1 - A_2)}  \lambda^{N_B} (1 - \lambda)^{N_{A_1} - N_{A_2}  - N_B} Z_{B + A_2} \\
    &\equiv \sum_{B  \subseteq (A_1 - A_2)} g_A(\lambda, N_B) Z_{B + A_2}.
    \label{eq:B2}
\end{aligned}
\end{equation}
With the definition Eq.~\eqref{eq:B2}, the SCEE can therefore be written as $S_s = - \ln \left( \frac{Z(\lambda = 1)}{Z(\lambda =0)} \right)$ and it can be further rewritten it to an integral expression $S_s = - \int_{0}^1 d \lambda \frac{\partial \ln Z(\lambda) }{ \partial \lambda}$. On the other hand, as the free energy $F$ of a canonical ensemble can be written as $F = - \frac{1}{\beta}\ln(Z)$, where $\beta \equiv \frac{1}{k_B T}$. The ratio of $Z(\lambda = 1)$ and $Z(\lambda = 0)$ can be expressed as, 
\begin{equation}
    e^{- \beta \Delta F} = \frac{Z(\lambda = 1)}{Z(\lambda =0)}.
    \label{eq:exp free energy difference and partition function ratio - appendix}
\end{equation}
By Jarzynski’s equality \cite{JarzynskiNonequilibrium1997}, $\langle e^{-\beta W_A} \rangle = e^{\beta \Delta F}$, we can rewrite the expression \eqref{eq:exp free energy difference and partition function ratio - appendix} as follows,
\begin{eqnarray}
    \langle e^{-\beta W_A} \rangle && = \frac{Z(\lambda = 1)}{Z(\lambda =0)},\nonumber \\
     -\ln \left( \langle e^{-\beta W_A} \rangle \right) &&= - \ln \left( \frac{Z(\lambda = 1)}{Z(\lambda =0)} \right), \nonumber\\
    S_s && = -\ln \left( \langle e^{-\beta W_A} \rangle \right),
    \label{eq:Subtracted EE relation with work done - appendix}
\end{eqnarray}
and the total work done $W_A$ in the tunneling process from $Z(\lambda = 0)$ to $Z(\lambda = 1)$ is,   
\begin{equation}
    -\beta W_A = \int_{0}^{1} dt \frac{d \lambda}{dt} \frac{\partial \ln g_A(\lambda(t)), N_B(t)}{\partial \lambda}  
    \label{eq:Work done integral expression - appendix}
\end{equation}
where $\frac{d\lambda}{dt}$ is set to 1 throughout this section.

\subsection{Empirical scaling study of quench time in non-equilibrium method}
\label{appB2}
As outlined in Appendix~\ref{appB1}, within the non‑equilibrium framework, the SCEE is related to the total work performed along the tunneling path from $Z(\lambda =0)$ to $Z(\lambda =1)$, an expression given in Eq.~\eqref{eq:Work done integral expression - appendix}. The number of discrete intervals used to evaluate this integral is referred to as the {\it quench steps}.

A significant practical limitation, however, is determining a quantitatively suitable number of quench steps for different system parameters and for different system sizes $L$. In prior works~\cite{zhaoMeasuring2022,liaoExtracting2024}, a fixed total number of steps $Q_t \sim 10^7$ was used for all $L$. This uniform choice is inefficient: for small $L$ it is unnecessarily large and wastes computational resources, while for large $L$ it may be insufficient, potentially compromising the reliability of the EE results. This indicates that the non‑equilibrium method, in its current form, remains incomplete.

To address this, we have performed an empirical scaling analysis of the required quench steps as a function of $L$. Our aim is to establish a benchmark $Q_t(L)$ that reduces the coefficient of variation (CV) to an acceptably small value for a given $L$. This approach refines the non‑equilibrium method, providing a systematic and resource‑efficient protocol, and places the technique on a more complete and robust footing.

Within the non‑equilibrium framework, the work $\beta W_A$ and the SCEE $S_s$ are related by $e^{-S_s} = \langle e^{-\beta W_A} \rangle$. We define the coefficient of variation of $e^{-\beta W_A}$ as the ratio of the sample standard deviation $\sigma_{e^{-\beta W_A}}$ to the sample mean $\mu_{e^{-\beta W_A}}$, i.e., $CV[e^{-\beta W_A}] = \sigma_{e^{-\beta W_A}} / |\mu_{e^{-\beta W_A}}|$.
Similarly, the CV of $\beta W_A$ is $CV[{-\beta W_A}] = \sigma_{{-\beta W_A}} / |\mu_{{-\beta W_A}}|$, where $\sigma_{-\beta W_A}$ and $\mu_{-\beta W_A}$ are the standard deviation and mean of $-\beta W_A$, respectively.

\begin{figure}[htp!]
	\centering
	\includegraphics[width=0.99\columnwidth]{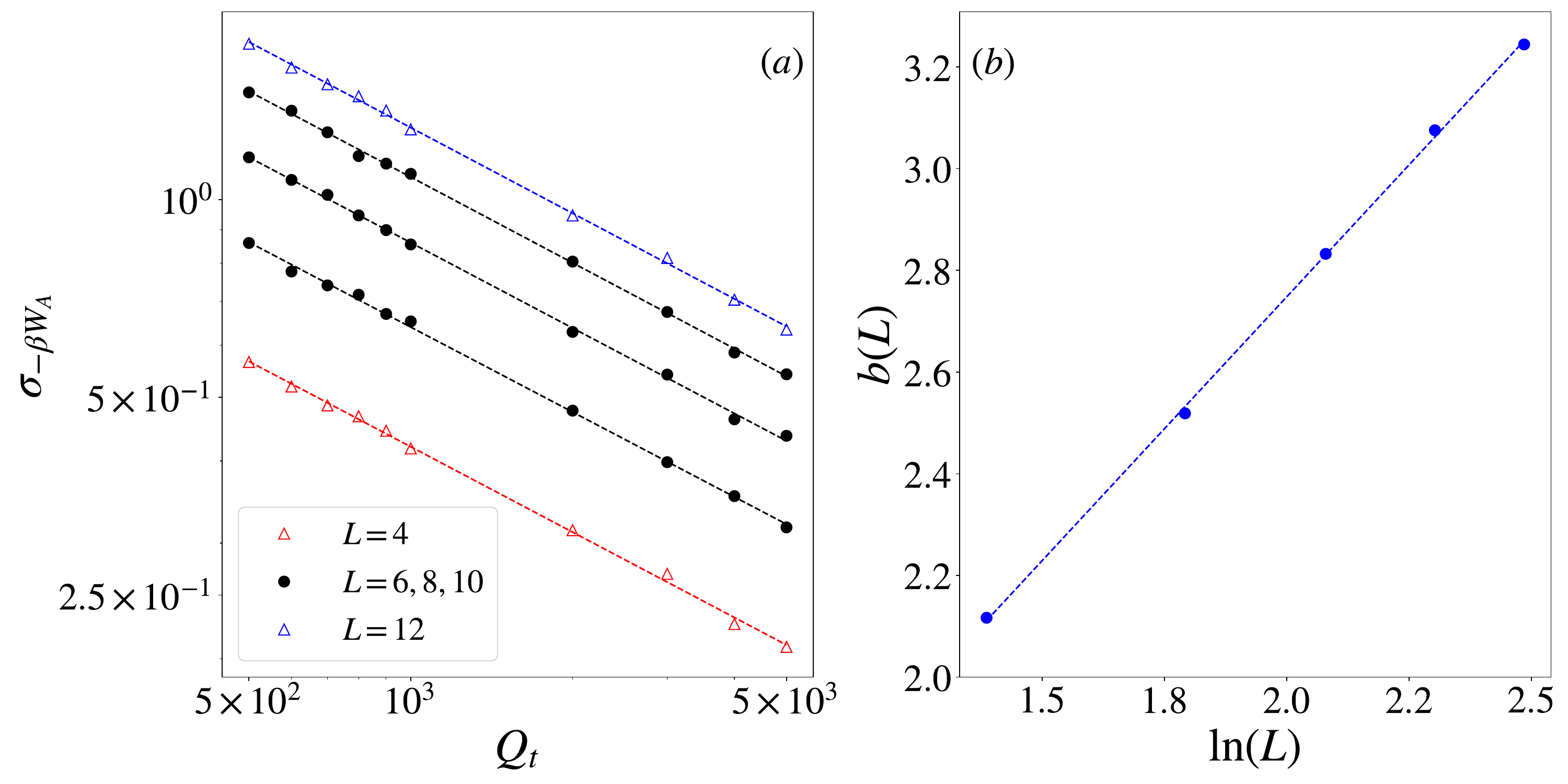}
	\caption{\textbf{Result of the empirical scaling analysis study}
        (a) The sample standard deviation {$\sigma_{-\beta W_A}$} against the total number of quench steps $Q_t$.  
        The average slope in (a) of the linear fitting $\bar{a} = -0.431(5)$.
        (b) The intercept $b(L)$ in (a) as a function of $L$. The slope obtained from linear fitting is $b_m=1.03(1)$, and the intercept is $b_0 = 0.67(2)$. 
    }
	\label{fig:Sample std versus qt}
\end{figure}

We find empirically that $\sigma_{-\beta W_A}$ exhibits a power‑law dependence on the quench step $Q_t$, as shown in the log–log plot of Fig.~\ref{fig:Sample std versus qt}(a). This leads to the empirical relation
\begin{equation}
    \ln(\sigma_{-\beta W_A}) =  \bar{a}  \ln Q_t +  b(L).
\end{equation}
Furthermore, as illustrated in Fig.~\ref{fig:Sample std versus qt}(b), the intercepts of the lines in panel (a) follow
\begin{eqnarray}
    b(L) = b_m \ln L + b_0.
\end{eqnarray}
Hence,
\begin{equation}
    \ln(\sigma_{-\beta W_A}) =  \bar{a}  \ln Q_t +  b_m \ln L + b_0.
\end{equation}

Mathematically, $\mu_{e^{-\beta W_A}} = e^{ \mu_{-\beta W_A} + \sigma^2_{- \beta W_A}/2}$ and $\sigma_{e^{-\beta W_A}} = \sqrt{ ( e^{\sigma_{-\beta W_A}^2} -1 ) \; e^{2 \mu_{- \beta W_A} + \sigma_{-\beta W_A}^2}}$. From this we obtain
\begin{eqnarray}
    &&CV[e^{-\beta W_A}] = \frac{\sqrt{( e^{\sigma_{-\beta W_A}^2} -1 ) \; e^{2 \mu_{- \beta W_A} + \sigma_{-\beta W_A}^2}}}{e^{ \mu_{-\beta W_A} + \sigma^2_{- \beta W_A}/2}} \nonumber \\
    &&=  \sqrt{e^{\sigma_{-\beta W_A}^2} -1} \nonumber\\
    &&=  \sqrt{(1 + \sigma_{-\beta W_A}^2 + \cdots ) -1} \nonumber \\
    && \approx \sigma_{-\beta W_A},
    \label{eq:Derivation of the CV and sigmaW relation}
\end{eqnarray}
provided $\sigma_{-\beta W_A}^2 \ll 1$. 

Consequently,
\begin{equation}
    Q_t(L) = \bigl( CV[e^{-\beta W_A}] \; e^{-b_0} \; L^{-b_m} \bigr)^{1/\bar{a}}.
\end{equation}
In principle, if we let $CV[e^{-\beta W_A}]$ decay with $L$ (so that the relative error decreases as $L$ grows), we can obtain a precise estimate of $e^{-S_s}$. 
In practice, we set $CV[e^{-\beta W_A}] = 0.05 / \sqrt{L}$ for all $L$. 
For example, as shown in Fig~\ref{fig:Sample std versus qt}, at $K=320$ on the phase boundary, we obtain $Q_t(K=320,L) = 4873 L^{3.56}$. 

\subsection{Benchmarking the non-equilibrium methods against PQMC result}
\label{appB3}
\begin{figure}[htp!]
	\centering
	\includegraphics[width=0.95\columnwidth]{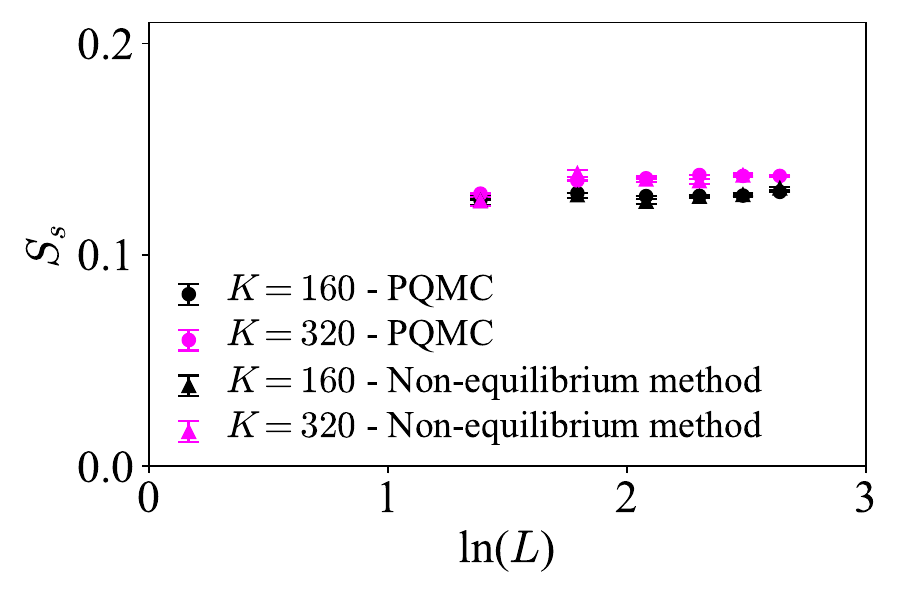}
	\caption{\textbf{Comparison of results from PQMC and the non-equilibrium method.} The SCEE $S_s$ is computed at the critical points $(K_c, h_c) = (160, 633)$ and $(K_c, h_c) = (320, 1260)$ using both methods. The simulations for $K=160$ and $K=320$ employ system sizes from $L= 4$ to $L = 14$.}
	\label{fig:Comparison of non-equilibrium method and PQMC result in larger K}
\end{figure}

To verify the accuracy of the PQMC results in the large-$K$ regime discussed in the main text, we extract the SCEE $S_s$ using a non-equilibrium method. Simulations are performed at the phase transition point $h_c$ for $K=160$ and $K=320$, corresponding to $h_c=633$ and $h_c=1260$, respectively. The results are presented in Fig.~\ref{fig:Comparison of non-equilibrium method and PQMC result in larger K}. As shown, the values obtained via the non-equilibrium method are in good agreement with those from the PQMC calculations.

\section{Measurement of Observables in the Bubble-Basis PQMC}
\label{appC}
As outlined in the main text, measuring an observable $O$ within the bubble basis can reduce to performing measurements in the $\sigma_z$ basis, since a bubble-basis state is an equal-weight superposition of multiple $\sigma_z$ basis states. Enumerating all these underlying $\sigma_z$ states for a measurement would be computationally prohibitive, as a single bubble corresponds to two distinct spin configurations. Therefore, a more efficient and practical strategy is to sample uniformly at random from the set of possible $\sigma_z$ states. In practice, for a given bubble-basis state, we generate 10 independent, uniformly random $\sigma_z$ configurations, compute $O$ for each, and take the average as a single measurement of $O$.

\begin{figure}[htp!]
\centering
\includegraphics[width=\columnwidth]{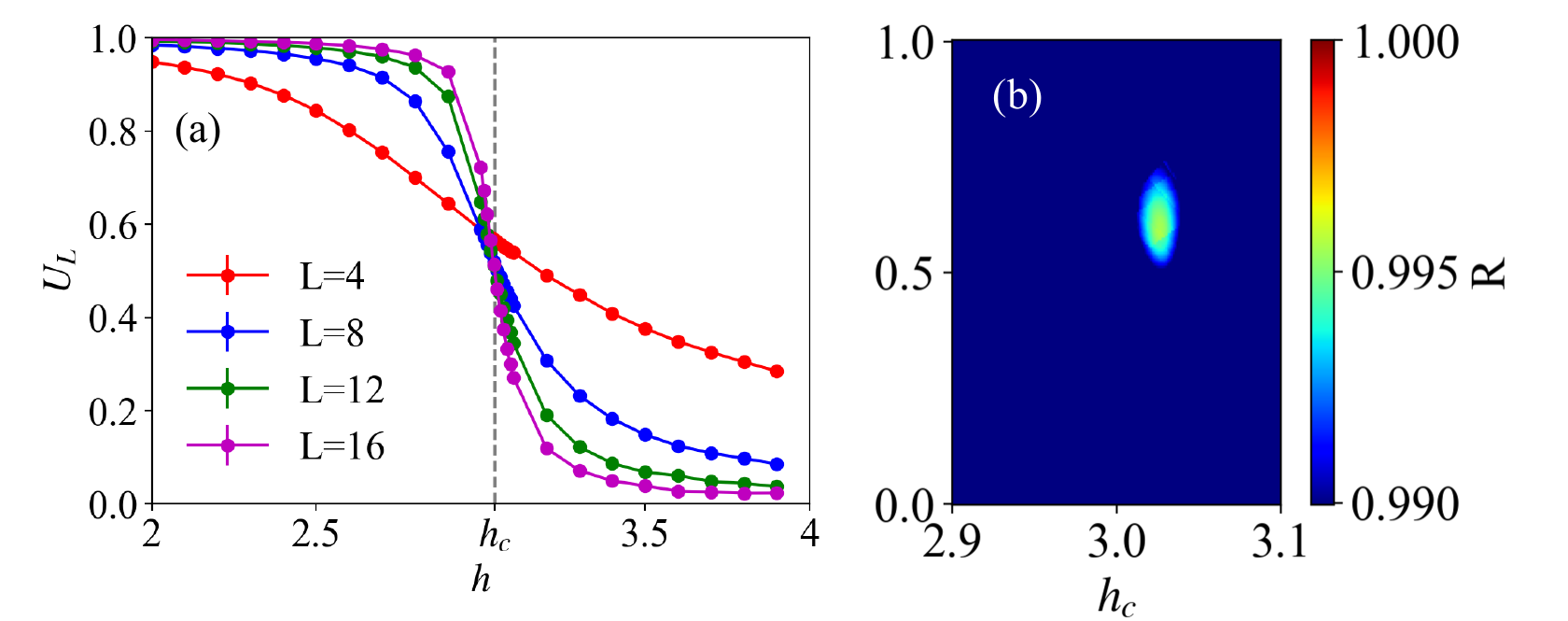}
\caption{\textbf{Finite-size scaling analysis of the (2+1)d Ising QCP sampled with bubble basis.} (a) Binder cumulant $U_L$ as a function of $h$ for different system sizes up to $L = 16$. The crossing point of the curves determines the critical field (b) Heat map of the $R$ statistic for the data collapse, evaluated over a range of $\nu$ and $h_c$ values. The optimal region of high $R$ aligns with $\nu = 0.63$.}
	\label{fig:bubble binder}
\end{figure}

To validate our bubble-basis implementation for the TFIM, we compute the Binder cumulant to locate the quantum critical point. For the (2+1)D TFIM, the established critical parameters are $h_c \approx 3.044$ and $\nu \approx 0.629$. Our results are shown in Fig.~\ref{fig:bubble binder}. Fig.~\ref{fig:bubble binder} (a) plots the Binder cumulant as a function of $h$ near the transition. The crossing point of the curves is at $h \approx 3.01$, marked alongside the established $h_c = 3.044$ (vertical grey dotted line), showing good agreement. Furthermore, Fig.~\ref{fig:bubble binder} (b) presents a heat map of the $R$-square statistic for data collapse across a range of $\nu$ and $h$ values. The optimal collapse yields estimates of $h_c \approx 3.01$ and $\nu \approx 0.62$, which are consistent with the known critical parameters for the (2+1)D Ising universality class~\cite{pfeutyIsing1971,hesselmannThermal2016}.
This consistency between our numerical estimates and the established critical exponents provides strong validation that the bubble-basis formulation correctly captures the critical physics of the TFIM.

\section{Implementation of Incremental Swap}
\label{appD}
As previously mentioned, the complexity of the incremental swap algorithm is $\mathcal{O}(mP)$. In this section, we detail the origin of this $\mathcal{O}(mP)$ scaling. A key factor is the efficient programming technique for recording and updating the bubble state with $\mathcal{O}(1)$ complexity. Furthermore, the specific steps for implementing Eq.~\eqref{eq:ratiofast} are also crucial and will be discussed in detail.

\subsection{Representation of the Bubbles}
\label{appD1}
To achieve the $\mathcal{O}(mP)$ time complexity, the choice of data structure for representing the bubble state is critical. We represent each state using a doubly linked list—a standard and widely useful data structure in computing. In this representation, each independent bubble corresponds to a connected component of nodes in the list. A single bubble is represented as a node whose left and right pointers both point to itself.

\begin{figure}[htp!]
\centering
\includegraphics[width=1\columnwidth]{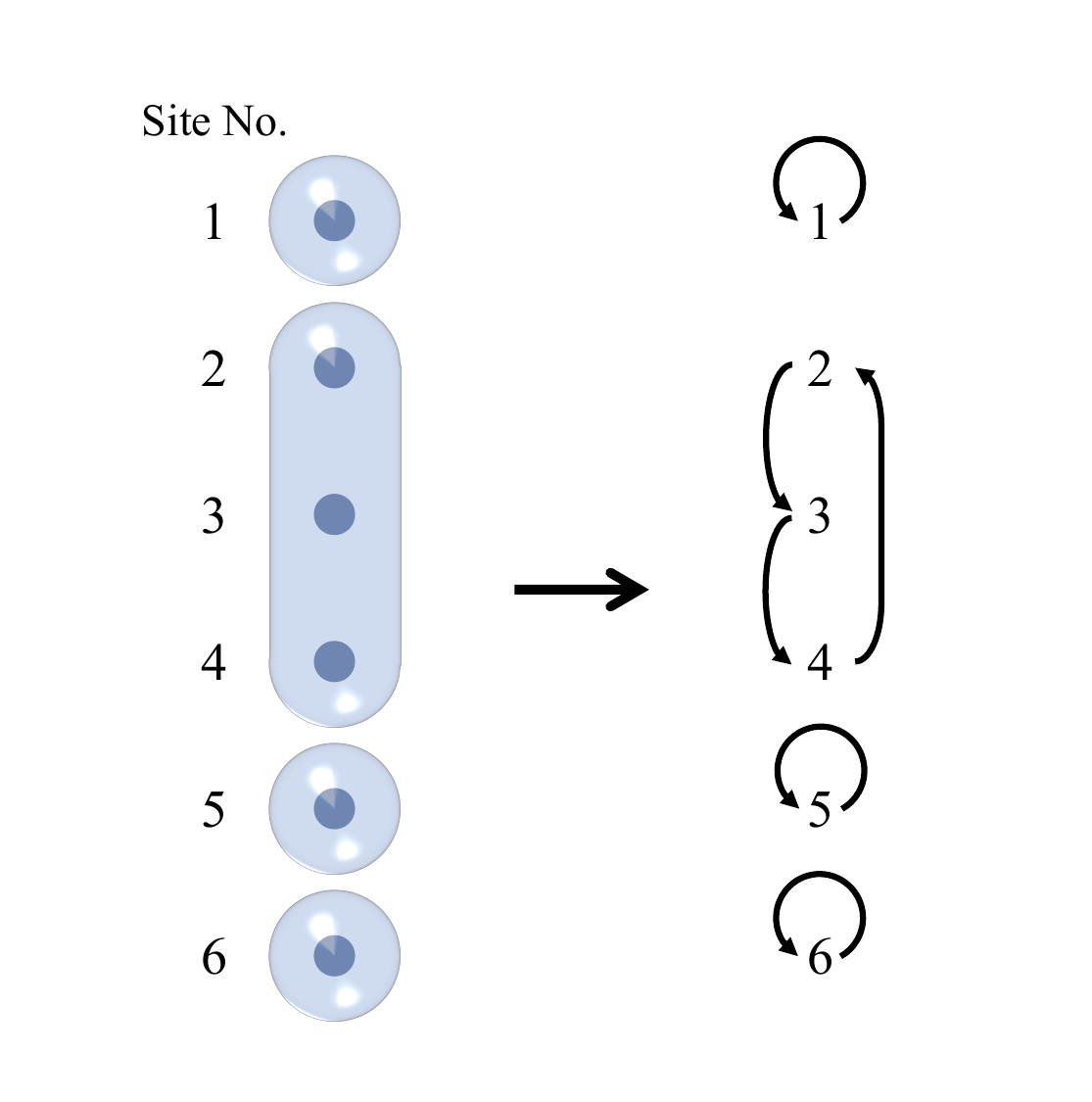}
\caption{\textbf{A figure demonstrates how the Bubbles were represented} The bubbles on the left represent Bubbles on the sites. A directed loop on the right depict a doubly linked list, with arrow tails indicating left pointers and arrow heads indicating right pointers.}
\label{fig:bubble represent}
\end{figure}

For example, as shown in Fig.~\ref{fig:bubble represent}, a bubble state with 6 sites and 4 independent bubbles can be represented as follows: for the first bubble, the right-pointer list is set as $\text{dlink}(1) = 1 $; for the second bubble, $\text{dlink}(2) = 3 $, $\text{dlink}(3) = 4 $, and $\text{dlink}(4) = 3 $; for the third bubble, $\text{dlink}(5) = 5 $; and for the fourth bubble, $\text{dlink}(6) = 6 $. The left-pointer list is simply the reverse of the right-pointer list.
This particular data representation is chosen because it significantly reduces the time complexity of merging, splitting, and reverting bubbles when applying an operator to a bubble state.

\subsection{Merging, Splitting and Revert a Bubble state}
\label{appD2}
When a quantum operator acts on a bubble state, it modifies the connectivity: the $H_J$ operator merges two distinct bubbles into one, whereas the $H_h$ operator splits a single bubble into two. The $H_K$ operator can be interpreted as two successive applications of $H_J$. Efficient implementations for merging and splitting are therefore essential. We describe these operations on our doubly-linked-list representation as follows.

\begin{figure*}[htp!]
\centering
\includegraphics[width=1.5\columnwidth]{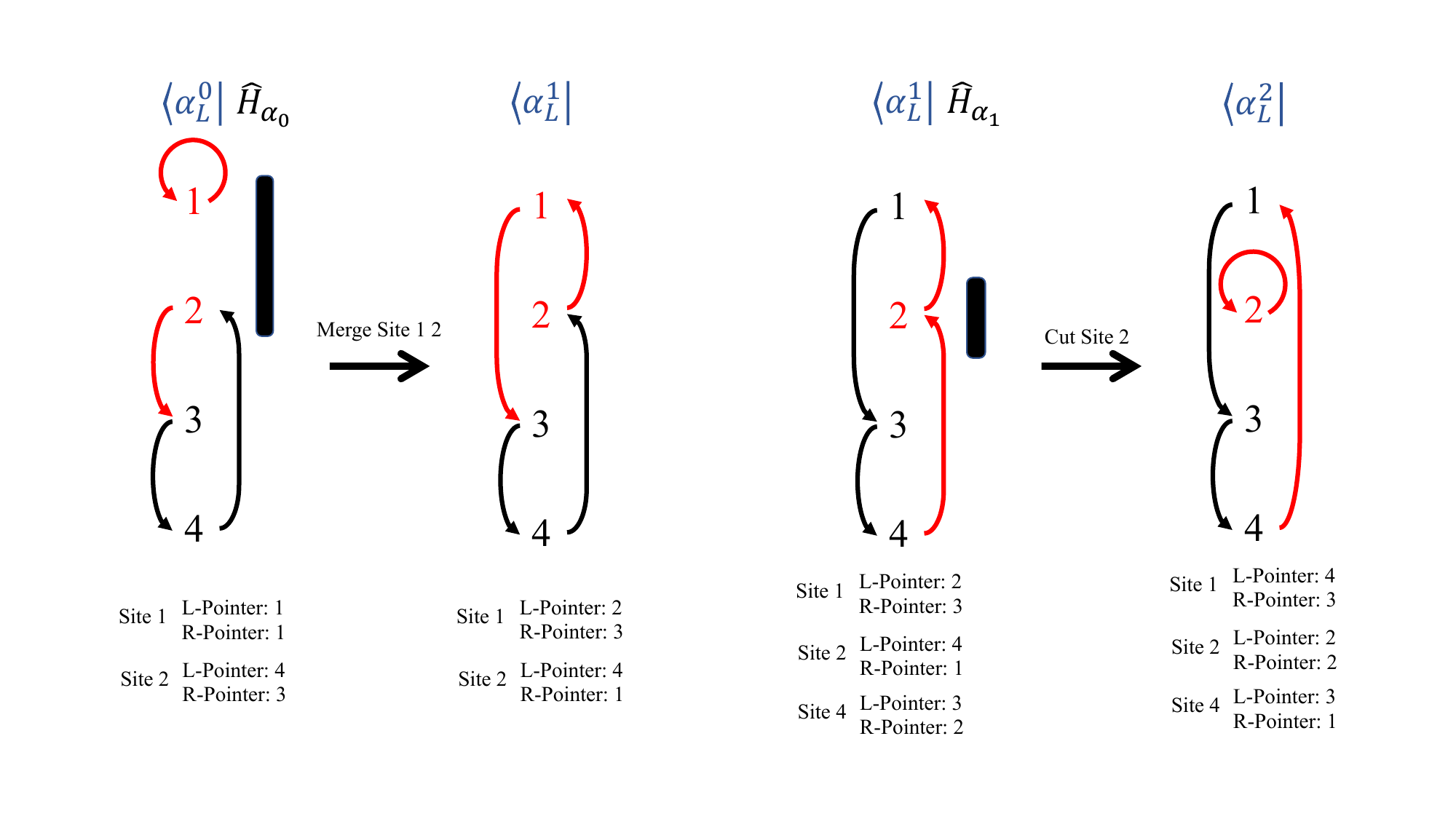}
\caption{\textbf{A figure demonstrates how the Bubbles were merged and cut} The left side shows how site 1 and 2 were merged together and the right side shows how the site 2 was separated from a large bubble. The changed node and pointers are labeled in red. The text underneath are showing the where are the left and right pointers pointing at of each node.}
\label{fig:bubble merge cut}
\end{figure*}

\begin{enumerate}
\item {\it Merging:} Suppose an operator merges Site 1 and Site 2. The right pointer of Site 2 is updated to point to Site 1, and the right pointer of Site 1 is updated to point to the node originally to the right of Site 2. Correspondingly, the left pointer of Site 1 is updated to point to Site 2. A diagram of this operation is shown on the left side of Fig.~\ref{fig:bubble merge cut}.

\item {\it Splitting:} Suppose an operator separates Site 2 from a larger bubble. We first reconfigure the pointers of its neighbors: the right pointer of Site 2's left neighbor is set to Site 2's original right neighbor, and the left pointer of Site 2's right neighbor is set to Site 2's original left neighbor. Subsequently, Site 2 is isolated by setting both its left and right pointers to point to itself, forming a new single-node bubble. A diagram of this operation is shown on the right side of Fig.~\ref{fig:bubble merge cut}.

\item {\it Revert:} To enable efficient backtracking during the Monte Carlo sampling, we store the original pointer values of all affected nodes before performing a merge or cut. Reverting an operation then simply requires restoring these saved pointers from memory.
\end{enumerate}

Each of these core operations—merge, cut, and revert—modifies only a constant number of pointers and therefore executes in $\mathcal{O}(1)$ time. Consequently, applying or reverting all operators along a single projector slice of length $m$ has an overall time complexity of $\mathcal{O}(m)$.

\subsection{Updating Overlap}
\label{appD3} 
\begin{figure}[htp!]
\centering
\includegraphics[width=1\columnwidth]{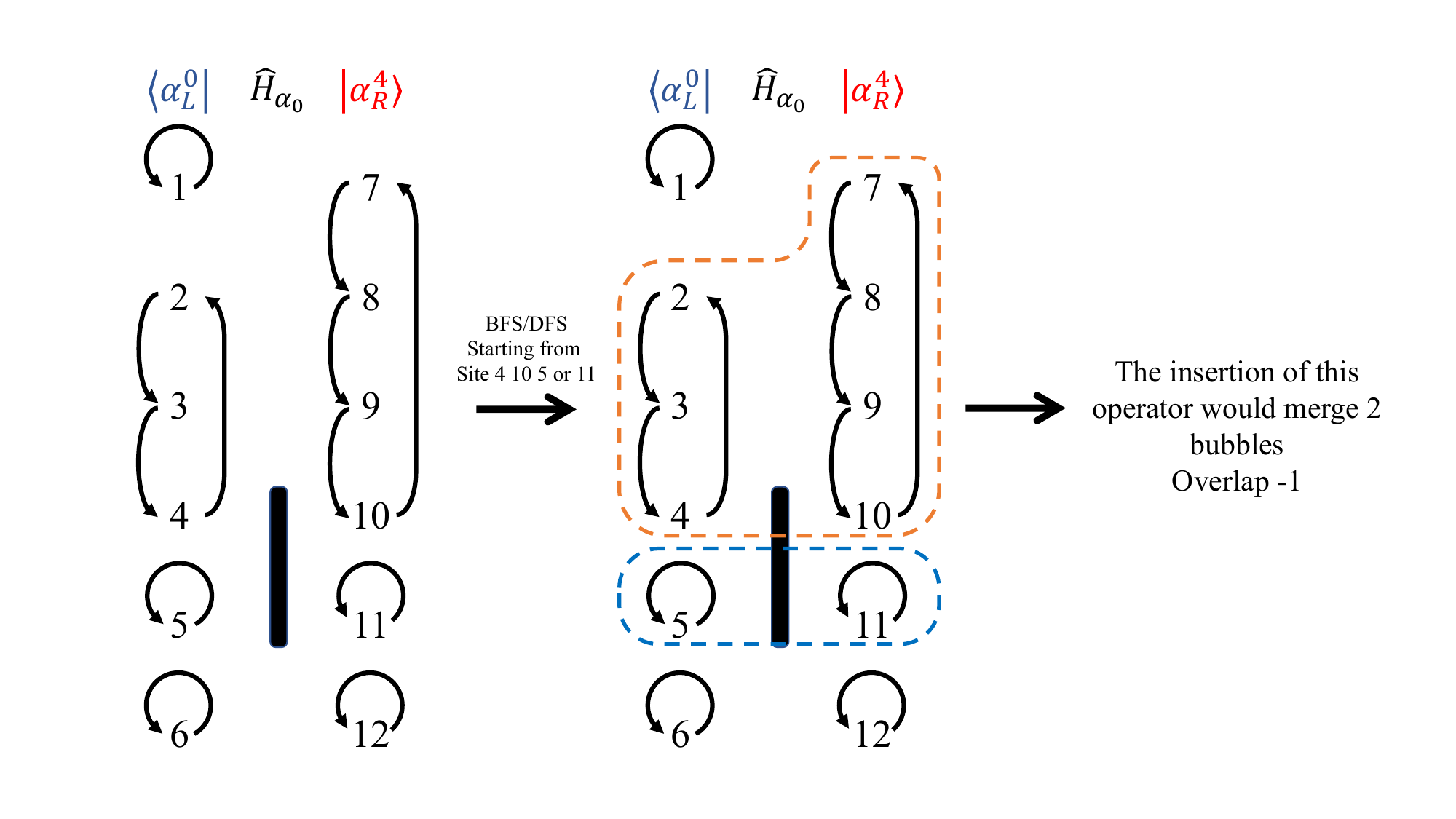}
\caption{\textbf{An example of inserting ${H_J}$ and resulting in overlap -1} As shown in the graph, the bubbles of site 4, 10 and 5, 11 are separate. Merging them resulting in overlap -1}
\label{fig:HJ_overlap_m1}
\end{figure}

\begin{figure}[htp!]
\centering
\includegraphics[width=1\columnwidth]{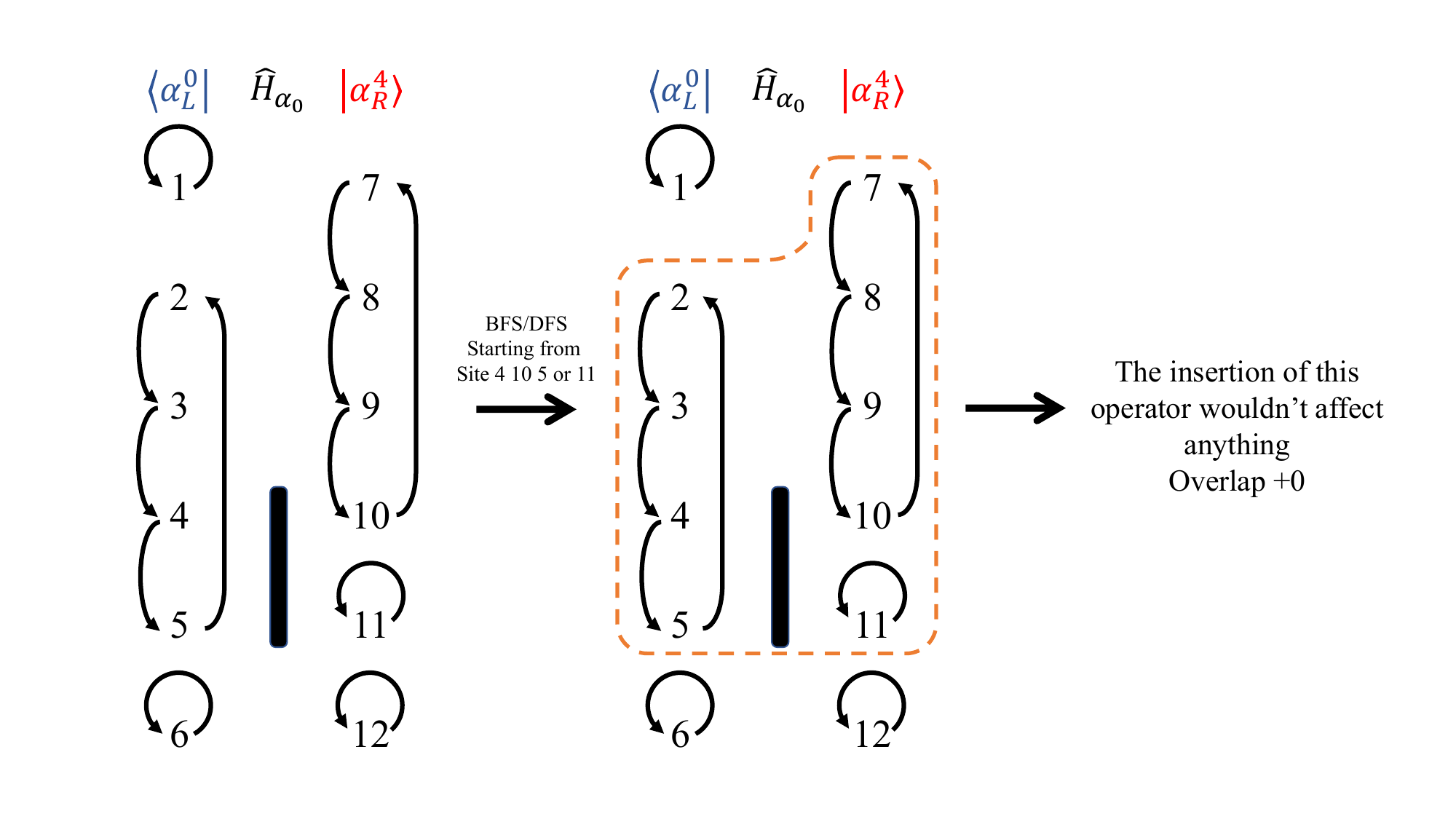}
\caption{\textbf{An example of inserting ${H_J}$ and resulting in overlap +0} As shown in the graph, the bubbles of site 4, 10 and 5, 11 are in the same bubble. Merging them affect nothing. Hence the overlap would not change}
\label{fig:HJ_overlap_p0}
\end{figure}

\begin{figure}[htp!]
\centering
\includegraphics[width=1\columnwidth]{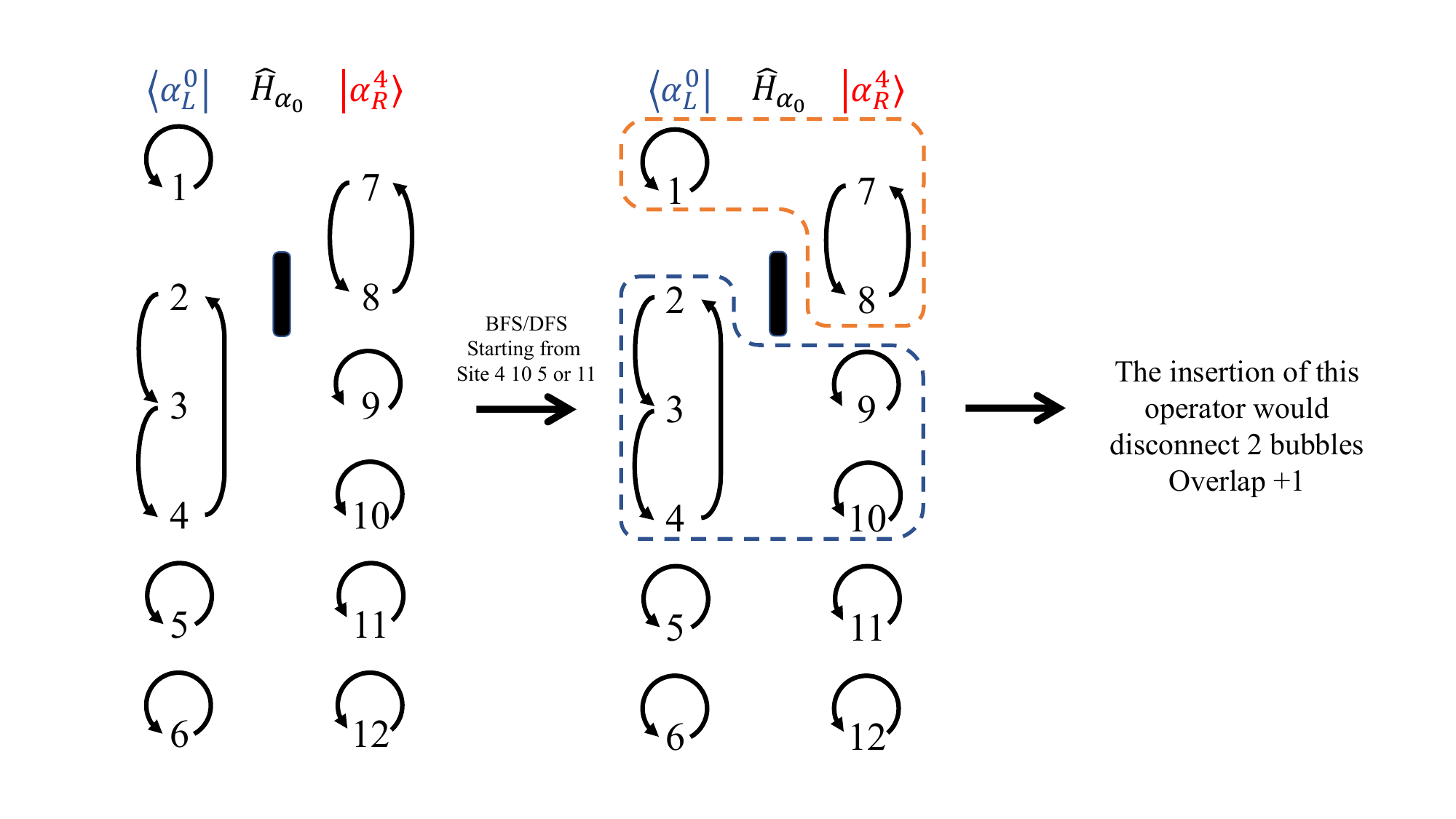}
\caption{\textbf{An example of inserting ${H_h}$ and resulting in overlap +1} As shown the graph, the insertion of this operator would seperate a bubble into 2. Therefore, overlap would be increased by 1.}
\label{fig:Hh_overlap_p1}
\end{figure}

\begin{figure}[htp!]
\centering
\includegraphics[width=1\columnwidth]{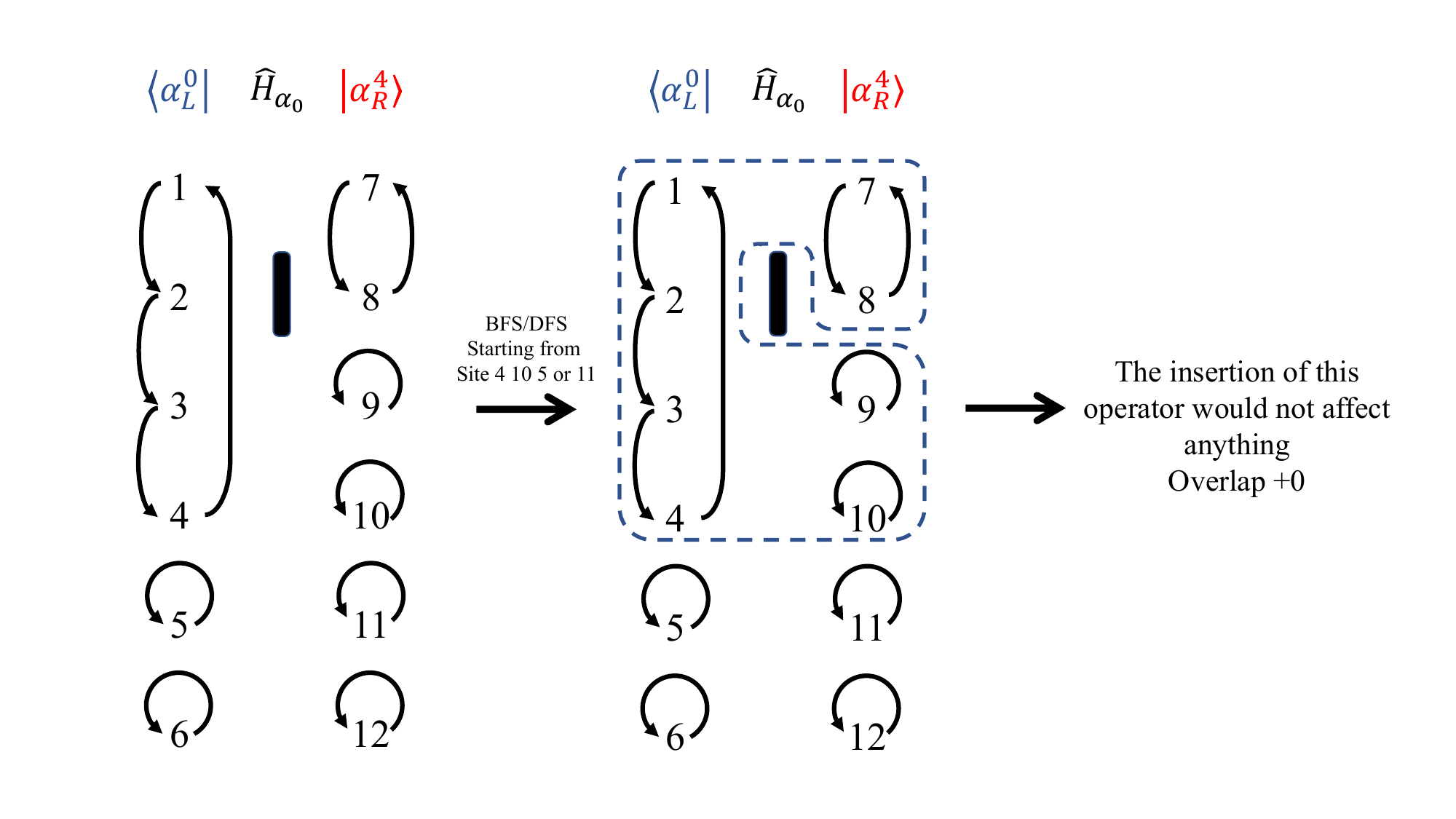}
\caption{\textbf{An example of inserting ${H_h}$ and resulting in overlap +0} As shown the graph, the insertion of this operator would affect nothing. Therefore, overlap would remain the same.}
\label{fig:Hh_overlap_p0}
\end{figure}

In this incremental swap algorithm, the overlap is updating for every projection slice based on the previous overlap of last slice.
For different operators the overlap updating strategies are different but are similar. The following are the details.
\begin{enumerate}
\item {\it Insert $H_J=(\sigma^z_i \sigma^z_j+\mathbf{I})$:} Check if site i and site j are in the same bubble or not by using breadth first search. If they are in the same bubble then the insertion would not affect the overlap. Otherwise, the overlap would decrease by 1. Since 2 bubbles are now 1 bubble. For graph of this operation and resulting in overlap -1, please refer to Fig.~\ref{fig:HJ_overlap_m1}. For resulting in overlap unchanged, please refer to Fig.~\ref{fig:HJ_overlap_p0}.
\item {\it Insert $H_h=(\sigma^{+}_{i}+\sigma^{-}_{i}+\mathbf{I})$:} Check if this operator when placing at site i would separate a bubble into 2 by breadth first search. To be more precise, see if site i is the only path that connects 2 bubbles. For graph of this operation and resulting in overlap +1, please refer to Fig.~\ref{fig:Hh_overlap_p1}. For resulting in overlap unchanged, please refer to Fig.~\ref{fig:Hh_overlap_p0}.
\item {\it Insert $H_K = (\sigma_a^z\sigma_d^z+\mathbf{I}) (\sigma_b^z\sigma_c^z+\mathbf{I})$:} Similar manner as inserting ${H_J}$. But checking 4 sites and 2 merging this time. By using breadth first search to classify which bubbles do site a,d,b and c belong to. After that, labeling site a,d,b and c by their bubble numbers. Count how many different labels are there. Merge the bubble of a,d and merge the bubbles b,c. Count how many different labels are there again. The overlap can then be calculated as the Number of different labels after minus the Number of different labels before.
\end{enumerate}
Most importantly, by using breadth first search for bubble connections, the complexity of all these indicating algorithm are $\mathcal{O}(P)$ on every projection slice. Which means the total complexity would be $\mathcal{O}(mP)$ for a complete MC sweep. This is the main reason why incremental swap can achieve $\mathcal{O}(mP)$.

\section{Convergence of Projection Length}
\label{appE}
\begin{figure}[h!]
    \centering
    \includegraphics[width=0.95\linewidth]{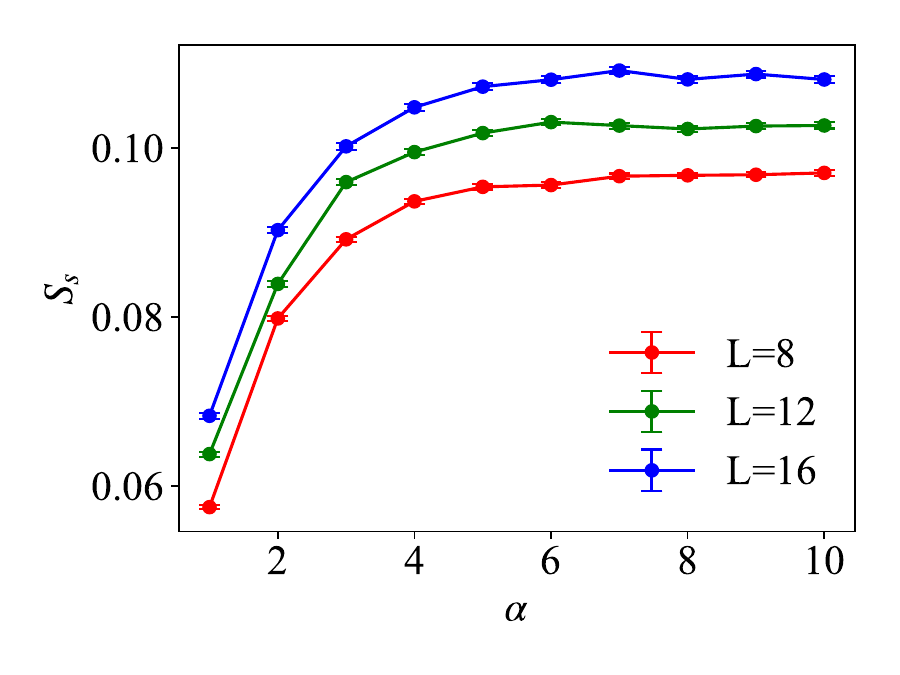}
    \caption{\textbf{Convergence of the SCEE $S_s$ with projection length $m$ at the Gaussian fixed point.} 
    We plot the SCEE $S_s$ as a function of the scaled projection length $\alpha = m/L^3$, calculated at the Gaussian fixed point for system sizes $L=8$, $12$, and $16$. Our results demonstrate that a projection length of $m = 8L^3$ is fully sufficient to ensure the obtained values of $S_s$ have converged to their asymptotic limit.}
    \label{fig:figs12}
\end{figure}

To justify our choice of projection length $m = 8L^3$ across the entire parameter space, we have performed convergence tests of the subtracted entanglement entropy $S_s$ as a function of the scaled projection length $\alpha = m/L^3$ at the Gaussian fixed point. The results are shown in Fig.~\ref{fig:figs12}.

Our results show that for the Gaussian fixed point at $L=8, 12$, and $16$, the value of $S_s$ is fully converged for $m = 8 L^3$. The guaranteed convergence of $S_s$ rigorously ensures that the universal corner coefficient $s$ obtained via linear fitting is also fully converged. The choice of $m=8L^3$ is further justified by the fact that the projection length required to converge to the ground state scales as $m \sim L^{d+z}$, where $d=2$ and $z=1$ is the dynamical critical exponent for both the Ising and Gaussian QCPs. The additional factor of 8 provides a conservative buffer to ensure full convergence across all parameter regimes.

\newpage
\bibliography{bibtex.bib}
\bibliographystyle{apsrev4-2}

\end{document}